\documentclass[amsmath,amssymb,ams, preprint, superscriptaddress,nofootinbib]{revtex4-2} %,twocolumn
\usepackage[a4paper,textwidth=17.9cm,textheight=26cm]{geometry}
\usepackage[english]{babel}
\usepackage{ulem}
\usepackage{siunitx} 

\addto\captionsenglish{}
\addto\captionsenglish{}

\usepackage{upgreek}
\usepackage{graphicx}

\renewcommand{\baselinestretch}{1.5} 

\usepackage[T1]{fontenc}
\usepackage{lmodern}
\usepackage{helvet}
\usepackage{lineno}
\usepackage{xr}
\makeatletter
\newcommand*{\addFileDependency}[1]{
  \typeout{(#1)}
  \@addtofilelist{#1}
  \IfFileExists{#1}{}{\typeout{No file #1.}}
}
\usepackage{pdfpages} 
\AtBeginDocument{\let\LS@rot\@undefined}
\makeatother

\begin{document}
\title{Large-area photonic circuits for terahertz detection and beam profiling}

\author{Alessandro Tomasino}
 \altaffiliation{Contributed equally to this work}
\affiliation{Hybrid Photonics Laboratory, École Polytechnique Fédérale de Lausanne (EPFL),  CH-1015, Switzerland}
 \affiliation{Center for Quantum Science and Engineering (EPFL), CH-1015, Switzerland}
 
\author{Amirhassan Shams-Ansari}
 \altaffiliation{Contributed equally to this work}
 \affiliation{Harvard John A. Paulson School of Engineering and Applied Sciences, Harvard University, Cambridge, MA, USA}
  \affiliation{DRS Daylight Solutions, 16465 Via Esprillo, San Diego, CA, USA}
\author{Marko Lončar}
 \affiliation{Harvard John A. Paulson School of Engineering and Applied Sciences, Harvard University, Cambridge, MA, USA}
 
\author{Ileana-Cristina Benea-Chelmus}
\affiliation{Hybrid Photonics Laboratory, École Polytechnique Fédérale de Lausanne (EPFL),  CH-1015, Switzerland}
 \affiliation{Center for Quantum Science and Engineering (EPFL), CH-1015, Switzerland}
\date{\today}

%\linenumbers

\begin{abstract}
Deployment of terahertz communication and spectroscopy systems relies on the availability of low-noise and fast detectors, with plug-and-play capabilities. However, most currently available technologies are stand-alone, discrete components, either slow or susceptible to temperature drifts. Moreover, phase-sensitive schemes are mainly based on bulk crystals and require tight beam focusing. Here, we demonstrate an integrated photonic architecture in thin-film lithium niobate that addresses these challenges by exploiting the electro-optic modulation induced by a terahertz signal onto an optical beam at telecom frequencies. Leveraging on the low optical losses provided by this platform, we integrate a double array of up to 18 terahertz antennas within a Mach-Zehnder interferometer, considerably extending the device collection area and boosting the interaction efficiency between the terahertz signal and the optical beam. We show that the double array coherently builds up the probe modulation through a mechanism of quasi-phase-matching, driven by a periodic terahertz near-field pattern, without physical inversion of the crystallographic domains. The array periodicity controls the detection bandwidth and its central frequency, while the large detection area ensures correct operation with diverse terahertz beam settings. Furthermore, we show that the antennas act as pixels that allow reconstruction of the terahertz beam profile impinging on the detector area. Our on-chip design in thin-film lithium niobate overcomes the detrimental effects of two-photon absorption and fixed phase-matching conditions, which have plagued previously explored electro-optic detection systems, especially in the telecom band, paving the way for more advanced on-chip terahertz systems.
\end{abstract}

\maketitle

\section*{Introduction}
The ever-growing demand for applications such as artifical intelligence, augmented reality, internet of things, wireless communication, and cloud-based computing is requiring ever-increasing bandwidths and frequency availability, pushing the next generation of communication systems (namely, 6G \cite{Dang2020What}) to utilize unprecedentedly high carrier frequencies, largely exceeding 100 GHz \cite{Koenig2013WirelessRate, Sengupta2018TerahertzSystems, Jiang2023PropagationBand}. The terahertz (THz) spectral range, with frequencies conventionally between 0.1 and 10 THz \cite{Ferguson2002MaterialsTechnology}, bridges the microwaves and optical domains and emerges as a valuable resource towards realizing these high-speed and high-fidelity communication channels \cite{Nagatsuma2016AdvancesPhotonics, Jiang2024TerahertzReview}. 
Compared to optical beams, THz radiation exhibits a much longer wavelength, thereby being much less affected by Mie scattering \cite{Drake1985MieScattering} and atmospheric scintillations (i.e., random fluctuations in the refractive index of the atmosphere \cite{Phillips1999TheoryScintillation}). This allows for the realization of communication channels that are more robust against environmental disturbances \cite{Mei2021EavesdroppingTurbulence, Wang2021SecrecySnow, Jing2018StudyAtmosphere}. On the other hand, compared to microwaves, free-propagating THz beams are significantly more directional, which could be of interest for point-to-point communication. However, the generation and detection of THz waves are often implemented through very high-gain antennas to compensate for the severe free-path loss occurring within the low atmosphere \cite{Yang2012UnderstandingAtmosphere}. Narrower beams allow for more accurate illuminations of the desired target, implying a more efficient use of the THz power and a reduced risk of eavesdropper attacks \cite{Shrestha2022JammingLink}. 
The higher directionality of THz waves demands developing a new set of hardware capable of keeping receivers oriented towards the THz communication link, especially in the case of non-stationary access points \cite{Ghasempour2020Single-shotNetworks, Guerboukha2024CurvingObstacles, Dong2022VersatileMultiplexing}. As such, the reliable and widespread adoption of THz technology is crucially reliant on the availability of versatile THz wireless detectors delivering high sensitivity for signals either collimated or focused \cite{Lewis2019ADetectors}. Current THz sensing technologies include bolometers \cite{Richards1994BolometersWaves, Cherednichenko2011ADetection, Shein2024FundamentalFrequencies}, graphene-based detectors (both for electronic and optical read-out) \cite{Ryzhii2013GrapheneBolometers, Cai2014SensitiveGraphene, Bandurin2018ResonantPlasmons, Lara-Avila2019TowardsGraphene}, Golay cells \cite{Golay1949TheDetector} and pyroelectric detectors \cite{Rashid1991PyroelectricApplications, Beerman1969TheRadiation}. These types of detectors are not easily compatible with available communication and sensing infrastructures due to practical limitations, such as the requirement of cryogenic operating temperatures, relatively high dark currents, and long recovery times. More importantly, their incoherent operational nature, i.e., being sensitive to the THz intensity rather than the THz electric field, results in the loss of the phase information. 

In contrast, coherent detectors are particularly important in communications and sensing, providing access to both the amplitude and phase of the terahertz electric field. Among these, the most common detection schemes are based on the coherent up-conversion of THz signals to the optical domain, either using photoconductive switches (i.e., photomixer) \cite{Auston1975PicosecondSilicon, Kono2001UltrabroadbandSampling} or the electro-optic effect in bulk crystals exhibiting a $\chi^{(2)}$ nonlinearity. In the latter case, the terahertz field induces an amplitude-dependent polarization modulation on a probe pulse via frequency up-conversion. These detection schemes operate at room temperature and most importantly provide very low-noise readouts~\cite{Benea-Chelmus2019ElectricState, Kutas2020TerahertzSensing}. 

However, commonly used crystals such as zinc telluride \cite{Wu1995Free-spaceBeams, Nahata1996CoherentSampling, Tomasino2013WidebandSampling, Wu2018TerahertzLaser}, gallium arsenide and gallium phosphide \cite{Wu19977Sensor} exhibit precise phase-matching wavelengths, due to their inherent dispersion relations in both the optical and terahertz domain (Fig.~\ref{fig1}a). A fixed operation wavelength severely restricts the choice of laser technologies that can be employed (e.g., Ti:Sapphire for zinc telluride). A further complication is that the phase-matching wavelength often overlaps with the range of two-photon absorption, resulting in severe non-linear absorption, and limiting the maximum probe power. In these free-space electro-optic detection schemes (Fig. \ref{fig1}b), the maximum modulation is achieved by strongly focusing the THz beam down to the diffraction limit. Owing to the up-conversion mechanism, the THz detection relies on detecting the polarization modulation via ellipsometry using commercially available optical detectors. However, since the THz spot is often much larger than the probe spot, the THz energy is typically not efficiently converted to the optical domain. 

A promising approach to control the phase-matching wavelength and simultaneously address shortcomings of bulk systems is to transition these technologies to on-chip~\cite{Chen2019AnSilicon, Messner2023PlasmonicModulators, Smajic2024PlasmonicReview}. For instance, integrated antennas allow for simultaneously targeting a specific terahertz frequency range where the performance is optimized and achieving strong field enhancement beyond diffraction limit\cite{Ibili2023ModelingRange}. This is in contrast with THz systems based on bulk crystals, where the sensitive bandwidth is dictated by the phase-matching condition. Consequently, THz antennas were integrated with chip-based plasmonic waveguides based on organic electro-optic molecules ~\cite{Salamin2019CompactDetector, Benea-Chelmus2020Electro-opticFrequencies, Ummethala2019THz-to-opticalModulator}. Despite the large second-order nonlinearity of the organic molecules, plasmonic approaches suffer from large propagation losses of ($\mathrm{0.25~dB/\mu m}$), which prevents the integration of multiple antennas in sequence, thus hindering the realization of large photonic circuits. Therefore, those devices were hampered by an inefficient collection of the THz power due to the large mismatch between the THz spot and the antenna collection area. 
Given these constraints, it is imperative to develop strategies that enable an improved collection and up-conversion of THz waves to maintain both a high signal-to-noise and a spectral selectivity to millimeter-sized incident THz waves. Furthermore, the silicon-based integrated circuits suffer from large two-photon absorption, thus limiting the on-chip optical probe power and resulting in poor signal-to-noise ratio of the electro-optic detection~\cite{Yin2007ImpactWaveguides}. 
Amongst available integrated photonics platforms, thin-film lithium niobate (TFLN) is particularly well suited, as it can achieve phase-matching outside the two-photon absorption range, for example at 1550~nm where fiber technologies are well developed~\cite{Herter2023TerahertzPlatform,Lampert2024Photonics-integratedLines,Zhang2024MonolithicGeneration}.
Here, we propose a low-loss electro-optic-based terahertz detector operating at a nominal frequency of 0.5 THz~(Fig.~\ref{fig1}c). We exploit the low propagation losses of 1.3 dB/m \cite{Zhu2024Twenty-nineNiobate} and absorption-limited loss of 0.2 dB/m \cite{Shams-Ansari2022ReducedWaveguides}, along with the high Pockels coefficient ($r_{33} \approx$ 30.9 pm/V \cite{Zhu2021IntegratedNiobate}) of lithium niobate (LN), to realize a millimeters-long Mach-Zehnder interferometer (MZI). The THz imparts a phase modulation onto the optical probe and is collected from an increased area defined by an array of 18 gold bow-tie antennas, each providing a 60-fold enhancement of the THz field. An appropriately chosen distance between individual antennas allows for quasi-phase-matching, which extends the nonlinear interaction between the THz and optical probe beams, resulting in a coherent build-up of the probe modulation across the entire array. Our TFLN platform provides combined merits of about 25-fold larger THz field enhancement, 100 times larger collection area, 3-5 times larger $\chi^{(2)}$ (for THz-optical nonlinear interactions), 30 times lower linear loss of the telecom and twice larger energy gap compared to routinely used bulk crystals (summarized in Fig.~\ref{fig1}d). Furthermore, the nanoscale design allows us to partially mitigate the approximately twice higher loss of lithium niobate compared to zinc telluride. Altogether, these properties enable peak-to-peak time-domain modulation efficiencies exceeding $\mathrm{\eta = 0.8\times 10^{-3}}$, obtained under the illumination of a test THz pulse with a field strength of approximately 4.5 V/cm. Second, the phased array provides spectral sensitivity around the operating frequency, with a minimum linewidth of 46 GHz. Third, we demonstrate that our detector maintains similar performance even when subjected to off-center and out-of-focus THz beams. Finally, the large area of the detector enables efficient mapping of the THz beam profile through a single one-dimensional (1D) scan. This capability opens avenues for applications in radar systems and target-locking mechanisms for moving objects. 

\section*{Results}
\subsection*{Geometry and properties of the large area terahertz detector}
A schematic of our Mach-Zehnder interferometer-based detector and an optical microscope image of its top view is depicted in Fig.~\ref{fig1}c and Fig.~\ref{fig1}e, respectively. A series of THz antennas is realized by patterning the metallic contacts on both sides of a waveguide fabricated on an x-cut TFLN wafer. Such antennas allow for coupling the free-space THz radiation onto the chip (fabrication details of our devices are provided in Methods, whereas the exact geometric dimensions are listed in Supplementary Note 1). An optical probe beam traveling inside the TFLN waveguide splits into two at the input y-splitter of the interferometer. Our design aligns the optical mode polarization to the crystallographic z-axis of the TFLN waveguide, thus exploiting the $r_{33}$ Pockels coefficient \cite{Bahadori2020FundamentalModulators} - the largest available in a waveguide geometry (see the optical mode simulations in Supplementary Note 1). The THz beam is incident orthogonally from the backside of the high-resistivity silicon substrate and simultaneously illuminates all bow-tie antennas, leading to their synchronous resonance. The incident THz radiation is polarized along the z-axis, thus maximizing the collection efficiency of the bow-tie antennas (Fig.~\ref{fig1}c).  After crossing the arms of the interferometer, the two probes are recombined at the output y-splitter and then sent to an infrared photodetector for acquisition. 

\subsection*{Quasi phase-matching with terahertz antenna arrays}
In our detector, both the individual antenna and array configuration are designed in order to maximize the cumulative modulation of the probe beam induced by the impinging THz signal. Specifically, each antenna features a pair of elongated metallic electrodes (with a length $L_{gap}$), which significantly prolongs the THz-probe interaction and consequently improves the efficiency of the nonlinear interaction, in contrast to more conventional designs \cite{Peng2024DirectNanowires, Alfihed2021CharacteristicsEmitters, Maraghechi2011ExperimentalFrequencies}. Figure~\ref{fig2}a the THz electric field established across the z-y plane of the antenna, which reveals a homogeneous field distribution along the gap length, with a 60-fold and 30-fold field enhancement near the gold electrodes and in the center of the gap, respectively. In this geometry, the probe beam propagates through the gap within a time interval $\tau_{gap} = L_{gap}n_g/c$, where $n_g$ is the optical group index and \textit{c} is the speed of light in vacuum. By choosing a gap length much shorter than a THz wavelength (e.g., $L_{gap} \approx  \lambda_T/10$), the probe beam will take a fraction of the duration of a complete THz cycle to cross the entire gap length, i.e. $\tau_{gap} \sim 1/(10f_{T})$, with $f_{T} = c/\lambda_T$ being the resonant frequency of the antenna. This ensures that the phase modulation imparted by each antenna builds up constructively along its entire gap \cite{Salamin2019CompactDetector, Benea-Chelmus2020Electro-opticFrequencies}, accounting for a contribution \(\Delta\phi(t) \propto r_{33}E^{ant}_{\text{THz}}(t)\), linearly dependent on the instantaneous THz electric near-field \(E^{ant}_{\text{THz}}(t)\). 
In a geometry where a series of antennas resonate synchronously, the probe beam encounters the n-\textit{th} antenna at the time instant $t_n$, thus experiencing a phase delay \(\Delta\phi_n \propto r_{33}E_{\text{THz}}(t_n)\). As such, the total phase modulation accumulated at the end of the series is simply the sum over all $N_{ant}$ contributions, i.e., \(\Delta\phi_{tot} = \sum_{n=1}^{N_{\text{ant}}}\Delta\phi_n\). If the time instants $t_n$ are all different, \(\Delta\phi_{tot}\) may be much lower than that of a single antenna. Therefore, in order to effectively benefit from the array configuration, it is crucial to engineer its periodicity to allow for a coherent adding-up of \(\Delta\phi_{tot}\) beyond the value of the single antenna element. To this end, we implemented a mechanism analogous to quasi-phase matching~\cite{Ma2007Narrow-bandCrystal, Kitaeva2009GenerationCrystals, Nawata2014SensitiveTemperature}, yet without the requirement of performing periodic poling of the nonlinear crystal (Fig.~\ref{fig2}b). Along each arm of the interferometer, we designed the distance between two consecutive antennas ($D_1$) in such a way that the oscillation period of the THz electric field within their gap matches the arrival time of the probe beam at each antenna. This type of synchronization leads the probe beam to acquire identical phase modulation contributions $\Delta\phi^+$ while crossing each antenna along the array. Ultimately, this results in a cumulative phase modulation equal to \(\Delta\phi^{U} = {N_{\text{ant}}}\Delta\phi^+\) for the upper arm. 

The coherent accumulation of the phase modulation contributions effectively extends the coherence length of the nonlinear interaction beyond the size of a single antenna element, thus boosting the detection sensitivity. The value \(D_1\) defines the specific operating frequency of the array, which we refer to as the phase-matching frequency \(f_{\text{PM}} = \frac{1}{\Delta t_{1}}\) of the array, with $\Delta t_{1} = \frac{n_gD_1}{c}$ the group delay of the optical probe between two antennas. To benefit from a push-pull effect where the lower and upper arm contribute equally to the total phase modulation, the array in the lower arm is displaced by a distance $D_2 = D_1/2$ (Fig.~\ref{fig2}a) with respect to the upper one. Such a displacement causes the lower probe to lag behind the upper one by a time delay \(\Delta t_{2} = \frac{\Delta t_{1}}{2} = \frac{1}{2f_{\text{PM}}}\). As a result, the probe beam crosses the antennas of the lower array during the negative half cycle of the THz near-field oscillations, resulting in a phase retardation contribution \(\Delta\phi^{-}\) of an opposite sign compared to the upper arm. Consequently, the lower probe will build up a total phase modulation with a reversed sign \(\Delta\phi^{D} = N_{\text{ant}}\Delta\phi^{-}=-N_{\text{ant}}\Delta\phi^+\). 
The two arms of the interferometer are designed with slightly different optical path lengths so that the built-in phase difference $\phi_B$ for the telecom probe is equal to $\mathrm{\pi/2}$. Under this condition, the interferometer operates in the so-called quadrature point, providing an unmodulated output intensity $I^{Q}_{out}$ that is half of that at the maximum transmission $I^{0}_{out}$ as indicated in Fig.~\ref{fig2}c. On this point, the curve slope ${dI_{out}}/ d\phi = I^{0}_{out}/2$ exhibits a maximum, thus allowing for the highest sensitivity to a phase change, and consequently to the incident THz electric field. Any additional phase imbalance $\Delta\phi^{U}$ (or $\Delta\phi^{D}$) due to the terahertz-induced Pockels effect, changes the output intensity by $\Delta I^{+}_{out}$ (or $\Delta I^{-}_{out}$). For small values of $\Delta \phi \ll \frac{\pi}{2}$, the intensity modulation is directly proportional to the phase modulation, i.e., $\Delta I_{out} \propto (\Delta\phi^{U}-\Delta\phi^{D}) \propto E_{THz}$, where $E_{THz}$ is the incident THz amplitude (see Supplementary Note 4).
In order to demonstrate the frequency selectivity and the quasi-phase matching mechanism of the interferometer, we derived an analytical expression describing the frequency response of our detector $\Delta I(f)$ (see Supplementary Information for a detailed derivation):
\begin{align}
        {\Delta I(f)} \propto \mathrm{\Delta\phi^{U}}-\mathrm{\Delta\phi^{D}} =  -2iE_{ant}(f){\frac{\sin ({\pi f N_{ant}\Delta t_1})}{\sin{(\pi f\Delta t_1})}\sin{(\pi f\Delta t_2)}}e^{i\pi f \Delta t_1(N_{ant}-1)}e^{i\pi f \Delta t_2}
    \label{array_freq_resp}
\end{align}
where \textit{f} is the THz frequency, $\Delta I(f)$ is the THz-induced intensity modulation of the probe beam at the output of the interferometer and $E_{ant}(f)$ is the THz electric near-field established inside the antenna gap, assuming a uniform THz illumination across the entire device. 
Equation \ref{array_freq_resp} provides guidelines to realize a detector sensitive at a desired frequency with a chosen bandwidth. In the specific case of $\Delta t_{2} = \Delta t_{1}/2 = \Delta t/2$,  Eq. \ref{array_freq_resp} becomes:
\begin{equation}
        \frac{\Delta I(f)}{E_{ant}(f)} \propto T_{MZI}(f) = -2i{\frac{\sin ({\pi f N_{ant}\Delta t})}{\sin{(\pi f\Delta t})}}\sin{\frac{\pi f\Delta t}{2}}e^{i\pi f \Delta t(N_{ant}-1/2)}
    \label{array_freq_resp_antisymmetric} 
\end{equation}
where we have introduced $T_{MZI}(f)$, defined as the complex spectral response of the device. In the Supplementary Information Note 4 we demonstrate that a higher number of antennas inversely decreases the detector bandwidth. At the same time, the peak response quadratically increases as more antennas are added to the interferometer arms. This outcome is in line with the quasi-phase-matching theory. 

To experimentally demonstrate the proposed quasi-phase-matching mechanism, we measured the time-domain response of our detector under the illumination with a collimated broadband coherent THz pulsed beam (Fig.~\ref{fig2}d). This allows for reconstructing its time- and frequency-domain response. To record the temporal response, we performed on-chip electro-optic sampling using femtosecond laser pulses as optical probes (full measurement details in the Methods and Supplementary Information Note 2). Here, we generated phase-locked terahertz pulses via a photoconductive antenna excited by a femtosecond near-infrared laser. We then acquired the intensity of the out-coupled probe beam by scanning the mutual delay between the arrival times of the THz and probe pulses in the chip using a mechanical delay line. This allows for the acquisition of a time-varying waveform reproducing the temporal response of our detector. We tested our devices under a collimated THz beam with a diameter W = 4 cm, propagating in free space for a distance of CL = 57 cm before reaching our chip, respecting the assumption underlying Eq.~\ref{array_freq_resp}. We further verified the performance of our device when illuminating just the antenna array in the upper arm (Fig.~\ref{fig2}e) or the arrays in both the upper and lower arm (Fig.~\ref{fig2}f), by covering half of the interferometer with a metallic blade (Fig.~\ref{fig2}d). We present the measured THz transients in units of absolute amplitude modulation $\Delta V/V$ of the electrical signal acquired from the photodiode to back-track the modulation efficiency. We observe that the amplitude of the signal retrieved for the fully illuminated double array (f) is twice larger than that of the singly-illuminated array (e), confirming the outcomes of our analytical model. Briefly, to reproduce the experimental temporal response of the device, we simulated the broadband THz electric near-field established in the antenna gap $E_{ant}(f)$ and included it into the analytical expression of Eq.~\ref{array_freq_resp_antisymmetric}. More details are provided in the Supplementary Note. This way, we experimentally confirmed that the fundamental frequency $f_{PM}$ is highly enhanced for the case of the fully illuminated double array, with a phase-matching frequency of 487~GHz and a linewidth of only 46 GHz, in excellent agreement with simulation results predicting 487~GHz and 48~GHz, respectively (see Fig.~\ref{fig2}e for comparison). Finally, we observe in the inset of Fig.~\ref{fig2}g, that the frequency component at $2f_{PM}$ is much stronger for the case of the single array, while the opposite occurs for the third harmonic at $3f_{PM}$, demonstrating clear suppression of even components in the double array case. 

These experimental findings demonstrate that our design has the ability to enhance the responsivity of a THz detector in a desired band spanning a few tens of GHz while suppressing its out-of-band response. This capability is beneficial for ensuring large data bandwidths and reducing cross-talk in communication systems. 

\subsection*{Operation under off-center illumination and beam profiling}
In many applications, another desirable feature for a detector is the capability of functioning under diverse illumination conditions and potentially performing an auto-correction of the device illumination to maintain optimum performance \cite{Li2023IntelligentVision}. Specifically, a THz detector should provide an adequate response even when the impinging THz wave is off-centered. To verify this capability, we studied the performance of our device as it moves away from the focal point. This is in contrast to the collimated illumination discussed in the previous section. We placed our chip into a focused THz spot and vertically shifted the device along the length of the interferometer (y-axis) as sketched in Fig~ \ref{fig3}a. The THz spot covers an area of around $660~\mathrm{\times}~700~\mathrm{\mu m^2}$, determined using a time-domain knife-edge measurement (see Supplementary Information Note 3), thus being smaller than the entire interferometer footprint. In Fig.~\ref{fig3}b, we show the THz waveforms acquired while moving the interferometer along the y-axis to align the THz spot with the various antennas. In all cases, we observe a multi-cycle THz waveform that oscillates with a period of roughly 2 ps and has an envelope maximum value that is independent of the exact location of the THz spot, providing similar dynamic ranges and hence robustness against misalignment. Moreover, we note that the peak of the envelope shifts from earlier towards later time delays as we move the chip upwards. In addition, the number of cycles within the envelope depends on the position of the THz spot, indicating an incomplete illumination for positions too far off the center. When the THz beam is centered on the most peripheral antenna pairs (the $1^{st}$ and $9^{th}$, respectively), roughly only half of the beam illuminates the array, resulting in an asymmetric envelope. However, as the THz spot moves towards more central positions along the array, a larger number of antennas are efficiently excited. Consequently, the recorded waveform becomes more symmetric with a longer duration of its envelope. The spectral amplitude at the $f_{PM} \approx ~ $0.5 THz component increases in value as the THz cross-section occupies more central positions, as visible from the power spectra in Fig.~\ref{fig3}c calculated via the Fourier-Transform of the curves in Fig.~\ref{fig3}b. Finally, we observe in Fig.~\ref{fig3}d that our device allows us to maintain a similar modulation peak frequency (within 6~GHz), regardless of which and how many antennas are illuminated. 

These observations provide insights into the temporal response and capabilities of our large-area detector. A modulation peak invariant with the position indicates that the time response of each antenna lasts only a few cycles, exhibiting a relatively fast decay. Because of this, when the probe beam crosses each antenna at the time instant corresponding to the peak of the THz near-field, the phase modulation due to previous antenna encounters is negligible. Therefore, we conclude that the instantaneous values of the reconstructed THz waveforms are always due to contributions from a single antenna. This effect is here achieved by choosing antennas with low quality factors and resonances significantly different in value from the phase-matching frequency ($f_{PM} = 487~$GHz and $f_{ant}=360~$GHz). This effect is especially apparent while comparing the THz transients acquired from central and peripheral illumination. Conversely, if the antennas had a relatively high quality factor (i.e., featuring long-lasting oscillations), the amplitude of the resulting waveform would keep growing with the number of antennas illuminated by the THz wave.

We took advantage of the fast response of the antennas to reconstruct the profile of the THz beam illuminating the chip, using the pairs of antennas as pixels. This is possible because the chosen spacing between antennas allows them to have collection areas that are not fully overlapped. Consequently, the near-field established within the gap is mainly dependent on the THz electric field locally captured by each antenna, and hence can be directly linked to the beam section just above the plane of the antenna. The latter is encoded in the amplitude of the various cycles composing the complete waveform, each of them uniquely associated to specific antenna pairs along the array. We now use the envelope of these waveforms to reconstruct a two-dimensional beam profile of the THz spot at the TFLN chip. As shown in Fig.~\ref{fig3}h, this is achieved by mapping the time axis of panel Figs.~\ref{fig3}f to a spatial position using the group velocity of the probe beam ($v_g = {c}/{n_g}$). The THz spot image reconstructed through our chip unveils a non-centered, elliptical shape. Cut-lines at $y = 0$ and $z = 0$ reported in the two insets of Fig.~\ref{fig3}h are accurately confirmed by beam profile measurements obtained via standard knife-edge measurements. This showcases the capability of TFLN circuits to be used for THz beam profiling, providing an adequate dynamic range even at significantly low THz field strengths.

\subsection*{Operation under out-of-focus terahertz beams}
Various terahertz applications in spectroscopy and communications typically require well-defined central frequency and bandwidth. Moreover, maintaining these characteristics across a wide range of illuminations is equally important. Here, we investigate to which extent the proposed quasi-phase-matching mechanism exhibits resilience to off-focused illumination. Specifically, we characterized a series of detectors operating at the same phase-matching frequency $f_{PM} = ~$ 487 GHz, yet featuring different numbers of antennas per array, i.e., 3, 6, and 9. This selection allows for direct control over the detection bandwidth using simple lithographic techniques, ensuring high reproducibility and robustness. The detector arrays were placed at three locations, i.e., 0, 5, and 15 mm away from the focal plane so as to be illuminated by a diverging THz beam, as sketched in Fig.~\ref{fig4}a. By applying Rayleigh’s law of diffraction for the THz wave, we calculated that the THz spot (around 0.7 mm) expanded to a diameter of 4.2 mm and 10 mm after the 5-mm and 15-mm propagating distances, respectively. This is sufficient to cover the entire array length on any device. Because of the reduced size of the 3-antennas detector (Fig.~\ref{fig4}b), we aligned its center with the THz spot and then kept the same illumination condition for the other two cases, as depicted in Fig.~\ref{fig4}f and Fig.~\ref{fig4}m. 

The THz transients recorded for each detector type are shown in Figs.~\ref{fig4}c,g,n for an increasing number of antennas, respectively. Since the arrays on the 3-antenna device are shorter than 1 mm across, the THz beam covers a great area of the detector already at the focal plane. As such, the corresponding waveforms do not significantly differ in terms of the number of cycles, whereas the transient amplitude decreases for larger longitudinal shifts along the THz path. This is due to THz beam diffraction resulting in a weaker THz electric field intensity. We note that in the frequency domain (Fig.~\ref{fig4}d) the spectral amplitude at $f_{PM}$ first increases and then decreases, as the THz beam diverges (Fig.~\ref{fig4}(d)-(h)-(p)). Here, both spectrum peak and linewidth associated with $f_{PM}$ are plotted as a function of the longitudinal shift (x-axis). While the THz modulation is the highest for intermediate THz beam sizes, the linewidth monotonically diminishes for increasing THz beam size covering the array. These results suggest that there is an optimum THz illumination condition for the operation of the device in terms of spectral response, as a result of two concurrent effects: field enhancement within the single antenna and additive contributions from a larger number of them. Finally, we notice a very similar behavior for the 6- and 9-antenna devices, as depicted in Figs.~\ref{fig4}g,h and Figs.~\ref{fig4}n,p, respectively, with some noticeable differences. The transient duration considerably changes for larger longitudinal shifts, especially for the 9-antenna case, since the initial THz illumination is only partially exciting all the antennas. This also leads to a faster decrease of the associated linewidth, which shrinks down to $\sim$40 GHz for the 9-antenna device, 15 mm away from the focal plane (Fig.~\ref{fig4}q). 

\section*{Discussion}
In summary, we demonstrated a coherent THz detector (i.e., measuring both phase and amplitude of the incident field) relying on the Pockels effect in TFLN photonic circuits, operating at a frequency of 500~GHz. Compared to state-of-the-art integrated terahertz detectors, the low optical losses of our platform enable the realization of a Mach-Zehnder interferometer hosting arrays of up to 9 THz antennas (on each arm) to effectively collect THz signals from free space. This is achieved by exploiting the considerably larger collection area provided by the array compared to a single antenna. Without suffering from optical losses, the optical probe beam can travel along a large circuit from TFLN waveguides and interact with the enhanced THz near-field generated at the gap of several antennas. 

Our THz-antenna-driven quasi-phase-matching mechanism ensures that all the collected fields contribute constructively at a selected THz frequency, thus effectively suppressing out-of-band sensitivity. This design is robust since the large and distributed detection area provides flexibility and high sensitivity under diverse types of THz beam illumination.
Compared to other phase-matching mechanisms, our work avoids complex periodic poling techniques ~\cite{Armstrong1962InteractionsDielectric,Yamada1993FirstorderGeneration} by quasi-phase-matching two electromagnetic waves of extremely distant spectral ranges, namely the THz and optical domains, via the lithographic patterning of field-enhancing THz antennas. Similar to quasi-phase matching, the periodicity of the array (analog to the poling period) and the number of antennas (analog to the length of the poled region) determine the center frequency and the detection bandwidth of the detector, respectively. 
With our approach, we demonstrated a relative bandwidth as narrow as $BW = \frac{FWHM}{f_{PM}} = 8.2\%$ that is highly relevant for applications sensitive to out-of-band channel jamming attacks~\cite{Shrestha2022JammingLink}. 

Furthermore, we showed that the detector can operate as a THz beam profiler, encoding the in-plane THz field into the time coordinate. This is enabled by the nearly single-cycle response of individual antennas, off-resonance to the phase-matching frequency of the device. This scheme could already pair with a feedback loop to open up possibilities for optimizing the illumination of the detector, similar to quadrant detectors, a missing component in the THz frequency range. While our proposal does not yet provide full two-dimensional and real-time terahertz imaging capabilities compared to state-of-the-art bolometric \cite{Oda2015MicrobolometerRegion,Zhou2024InnovationsDetection,Ajayan2024RecentReview} or field-effect \cite{Ma2024IntegratedDetection,Bodrov2021Terahertz-field-inducedField,Li2023SubcycleWaveforms} cameras, the sensitivity of our design to low THz field strengths, its linearity, and high dynamic range, and our proposed read-out methods are anticipated to sparkle future work in building more complex architectures such as a photonics-based THz camera by combining several devices in parallel with synchronous read-out. 

All the functionalities shown here represent a significant step towards plug-and-play and deployable THz field-resolved detectors that can be adopted in practical scenarios. By further engineering both the optical waveguides and the THz antennas, control over the device spectral response can be achieved to enable simultaneous operation on multiple THz bands. Alternatively, the single-period array may be replaced by a collection of frequency-chirped antennas, properly spaced, to achieve broadband detection. Furthermore, the entire array could be designed to be sensitive to THz radiation propagating parallel to its axis, rather than perpendicularly (i.e. an end-fire array type \cite{Kossey2018End-fireSpacing,Abdullah2020VivaldiPhotomixers}), paving the way for detection of THz waves locally generated on the same chip towards spectroscopy applications~\cite{Herter2023TerahertzPlatform}. Our detector requires relatively low optical energies of around 10 pJ and thus can be readily integrated with modern chip-scale femtosecond sources~\cite{Yu2022IntegratedNiobate, Yang2024Titanium:sapphire-on-insulatorAmplifiers}. By integrating well-established telecommunication photodiodes on the same device can enable a chip-scale THz detector, paving the way toward the realization of fully integrated THz spectroscopy systems, time of flight measurements, and THz communications, all in a single and portable miniaturized device. 

\begin{figure}[htb]
 \includegraphics[width=16.9cm]{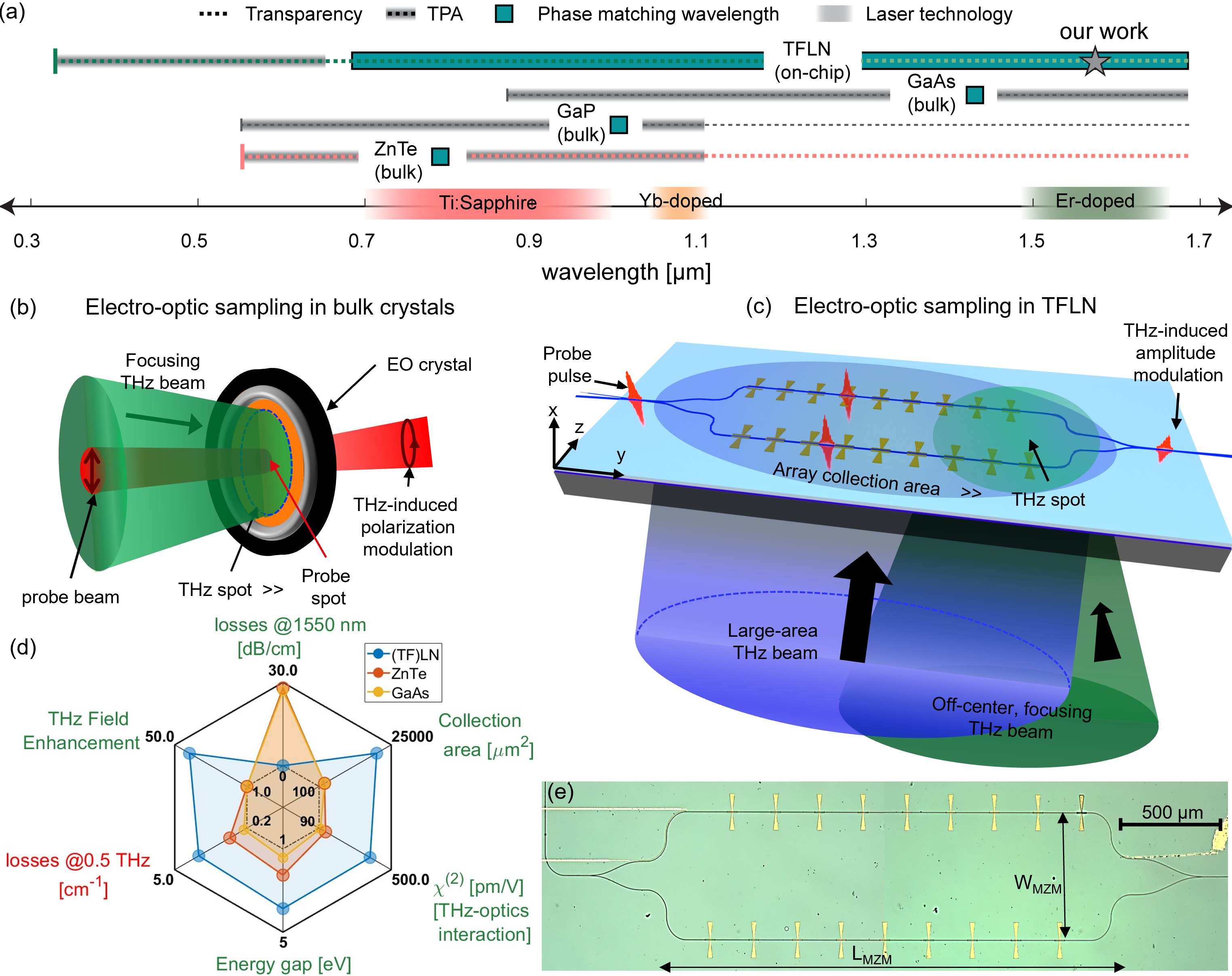}\renewcommand{\baselinestretch}{1}
   \caption{\textbf{Electro-optic sampling in bulk crystals versus on-chip in thin-film lithium niobate.} \textbf{(a)} Available material platforms for THz transient detection using the EO-sampling technique. Due to their inherent dispersion properties, the phase-matching wavelength is fixed for bulk crystals, such as ZnTe, GaAs, and GaP. The bottom axis shows the most widespread laser technologies used for EO-sampling, which do not always provide emission at the exact phase-matched wavelengths (see GaAs). Thin-film lithium niobate (TFLN) provides means to engineer group and phase velocities at both THz and optical frequencies, enabling phase-matching in a rather continuous optical range. Moreover, thanks to its large bandgap, TFLN allows for THz detection far from wavelength ranges affected by two-photon absorption (TPA). This is especially advantageous in the telecommunication band, where fiber optic infrastructure is available, whereas crystals such as GaAs are impractical due to TPA. \textbf{(b)} Conventional implementation of EO-sampling in bulk crystals. Both the optical and the THz beams are tightly focused onto the EO-crystal. The THz beam induces a field-dependent polarization modulation on the probe beam. The latter is measured via ellipsometry (not shown in the figure). In this approach, a great limitation to the THz signal-optical probe interaction is the large mismatch between the THz and optical spot sizes. \textbf{(c)} Implementation of EO-sampling in waveguide-based TFLN platform. This approach relies on the interaction of the probe pulse with an array of THz antennas deposited on each arm of an interferometer on-chip. The incident THz beam generates a phase modulation on the probe beam traveling through each arm, leading to the THz-induced amplitude modulation of the probe beam at the output of the interferometer. The antenna arrays effectively increase the sensitive area of the detector to a physical footprint of a few millimeters, extending the use of this type of device to operate with both large area and off-center THz beams. \textbf{(d)} Radar chart reporting a comparison among the main THz-detection platform properties for the cases of TFLN (blue), ZnTe (orange), and GaAs (yellow). Values represented by circles are data taken from \cite{Wu19977Sensor} and \cite{Tripathi2013AccurateSpectroscopy} for ZnTe, from \cite{Palik1997GalliumGaAs} and \cite{Nagai2004GenerationPulses} for GaAs, and from \cite{Zhu2024Twenty-nineNiobate} and \cite{Boyd1973MicrowaveNonlinearities} for TFLN. Numbers on the chart mark the lower and upper boundaries of each property range. TFLN supersedes the other platforms in THz field enhancement, TPA, optical losses, sensitive area, and nonlinearity. Terahertz losses are greater for bulk LN, yet TFLN technology mitigates this effect owing to the deep sub-wavelength size of the waveguides compared to the THz wavelength. \textbf{(e)} Optical micrograph of the Mach Zehnder interferometer realized in TFLN, with a length of $L_{MZM} = 2.6$ mm and a width of $W_{MZM} = 670~\mu m$.}
  \label{fig1}
   \vspace{-100pt} 
\end{figure}

\begin{figure}[htb]
  \includegraphics[width=17.5cm]{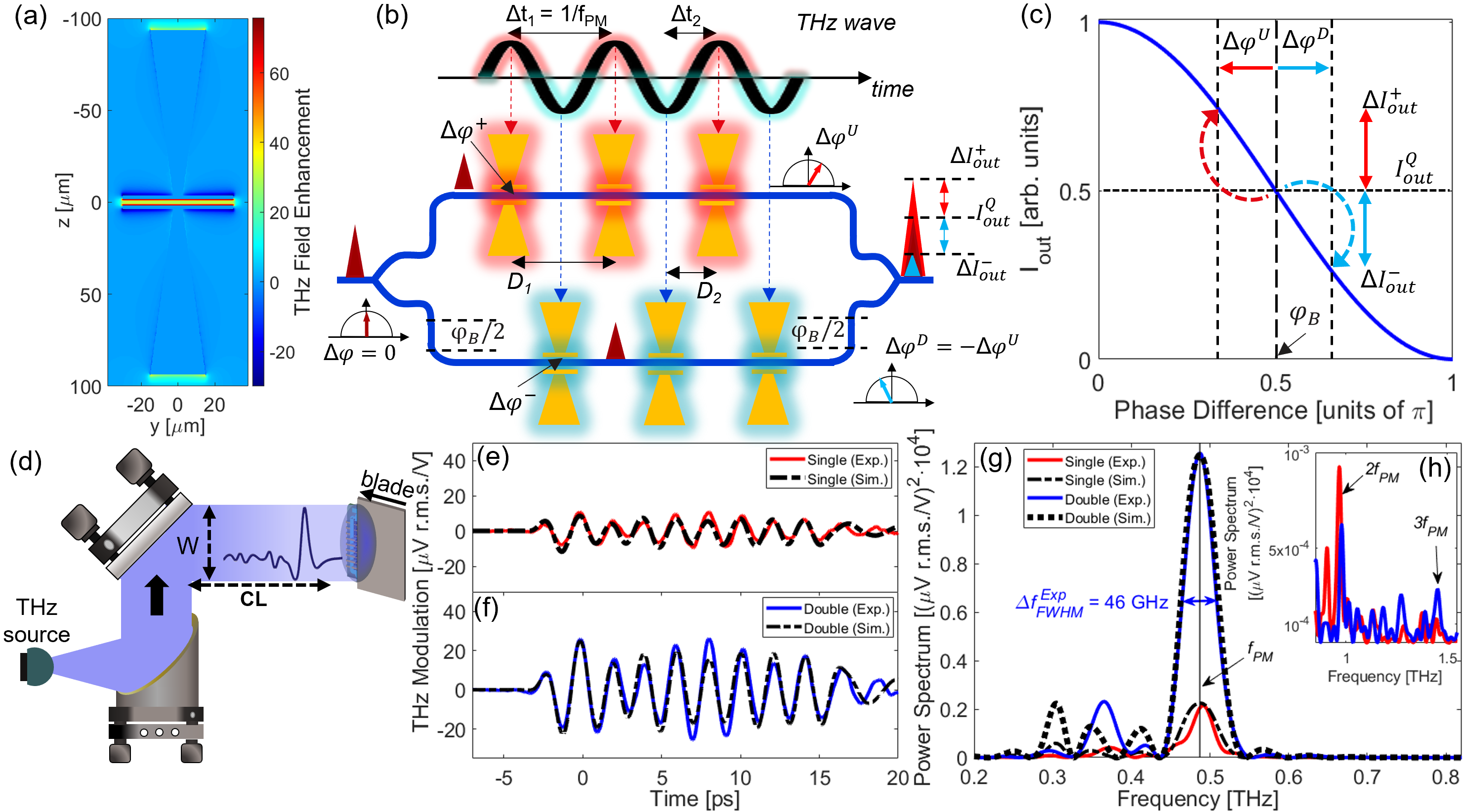}\renewcommand{\baselinestretch}{1} 
  \caption{\textbf{Quasi-phase-matching with antenna arrays.} \textbf{(a)} CST simulation of the enhanced THz electric field on the z-y plane of the antenna, at the resonance. The field is uniformly distributed along the entire gap ensuring coherent build-up of the probe phase modulation. \textbf{(b)} Sketch depicting the coherent build-up of the phase modulation imparted to the probe beam by the THz wave and leading to the amplitude modulation operated by the Mach Zehnder interferometer (MZI). The MZI is realized with an arm length difference leading to a build-in phase imbalance ($\phi_B$) allowing for operation at its quadrature point. A probe beam with with an initially null THz-induced phase retardation ($\Delta \phi = 0$) enters the MZI and is split into two identical beams. Each probe crosses an array of $N_{ant}$ antennas (shown in number of 3 for simplicity), where it experiences a phase retardation that is proportional to the THz electric field established in the antenna gap. The spatial period of the array ($D_1$) sets the arrival of the probe beam at each antenna at multiples of the time interval $\Delta t_1$. This leads to a coherent build-up of all phase modulation contributions imparted along the entire array. If the lower array is displaced by a distance $D_2$ (corresponding to a time interval $\Delta t_2$), the probe beam in the lower arm will cross each antenna when the THz field oscillations exhibit an opposite polarity compared to the top arm. The latter will impart a total phase modulation of a reverse sign, interfering with that of the top arm and leading to the intensity modulation of the probe beam at the output of the MZI. \textbf{(c)} Transmission curve of the interferometer operating at the quadrature point excited by a bipolar THz wave where the phase modulation $\Delta \phi^U/\Delta \phi^D$ changes the output probe intensity by $\Delta I_{out}^+/\Delta I_{out}^-$. Scanning the mutual delay between THz and probe beams allows for recording the time evolution of the THz field transient into the array after excitation of the incident THz pulse from the free space. \textbf{(d)} Experimental configuration to demonstrate the phase-matching mechanism. The THz beam is collimated with a diameter of W = 4 cm and sent to the device after propagating for CL = 57 cm in free space. To test the case of the single-arm illumination, half of the interferometer is covered with a metallic blade. \textbf{(e)} THz electric field waveforms reconstructed for the case of single arm illumination (red solid line) and \textbf{(f)} double arm illumination (blue solid line). In both panels the black dash/dotted lines represent the results of the analytical model computed via Eq.~\ref{array_freq_resp}, using the simulated complex field $E_{ant}(f)$ as described in the main text. \textbf{(g)} Power spectra obtained by Fourier transform of the waveforms in (e) and (f), for both experimental and simulated cases. The full-width half-maximum linewidth $\Delta f^{Exp}_{FWHM}$ retrieved for the double arm illumination shows an excellent agreement with that calculated analytically and shown in panel (g). \textbf{(h)} zoom-out of plot in (g) showing higher harmonics.}
  \label{fig2}
\end{figure}

\begin{figure}[htb]
    \centering
    \includegraphics[width=17.5cm]{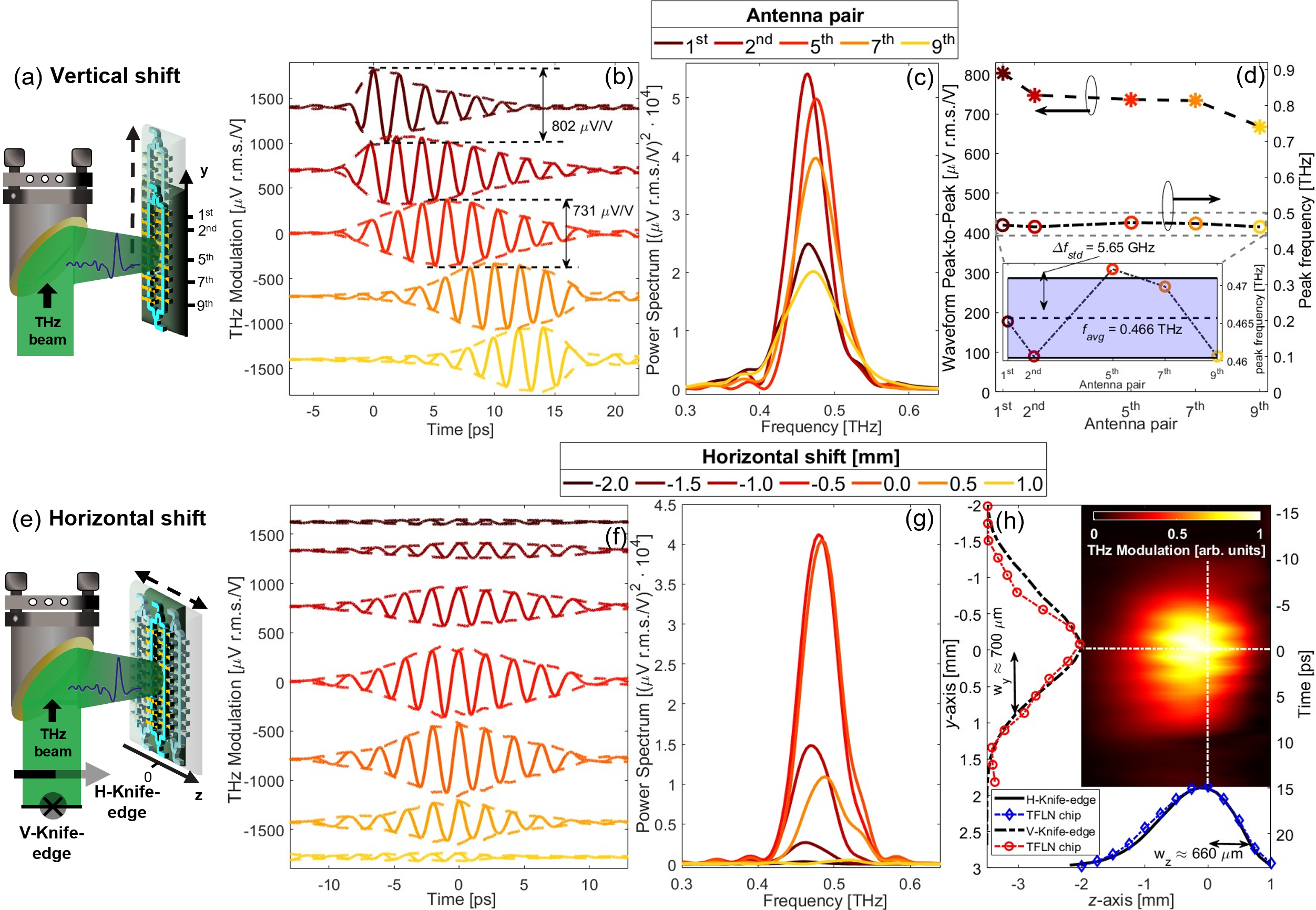}
    \renewcommand{\baselinestretch}{1} 
    \caption{\textbf{Large area TFLN circuits for waveform detection and beam profiling in the focal plane.} \textbf{(a)} Experimental configuration for measuring the response of the TFLN chip at different positions of the THz spot along the antenna array (y-axis). \textbf{(b)} Terahertz waveforms recorded for various values of vertical shifts (solid lines) and fitted envelopes (dashed lines). The position of the peak of the envelope coincides with the pair of antennas that are spatially aligned with the THz beam. The maximum THz modulation remains relatively constant across all measurements. All curves are vertically separated for clarity. \textbf{(c)} Calculated power spectra of the waveforms in (b), showing larger spectral amplitudes at $f_{PM}$ for the case where the THz spot is aligned with the center of the antenna array. \textbf{(d)} Terahertz modulation peak (stars) and peak frequency (open circle) show little dependence on any specific antenna pairs under illumination. A zoom-in of peak frequency behavior upon horizontal shifts is given in the inset in (d), showing that the operating frequency only varies by a few percent of the average value. \textbf{(e)} Experimental configuration for measuring the TFLN chip response as a function of the horizontal shift (z-axis) and to reconstruct the beam profile at the chip. \textbf{(f)} Terahertz waveforms recorded for various values of horizontal shifts (solid lines) and fitted envelopes (dashed lines). The position of the peak of the envelope and the maximum THz modulation remains relatively constant similar to (b). Curves are vertically separated for clarity. \textbf{(g)} Calculated power spectra of the waveforms in (f). \textbf{(h)} Two-dimensional image (z-y plane) of the focused terahertz spot reconstructed combining the waveforms shown in (f). Cut-lines at z = 0 (red circles) and y = 0 (blue diamonds) overlaid with curves profiles (black dotted/dashed line) retrieved via knife-edge measurements.}    
  \label{fig3}
\end{figure}

\begin{figure}[htb]\centering
 \includegraphics[width=17.5cm]{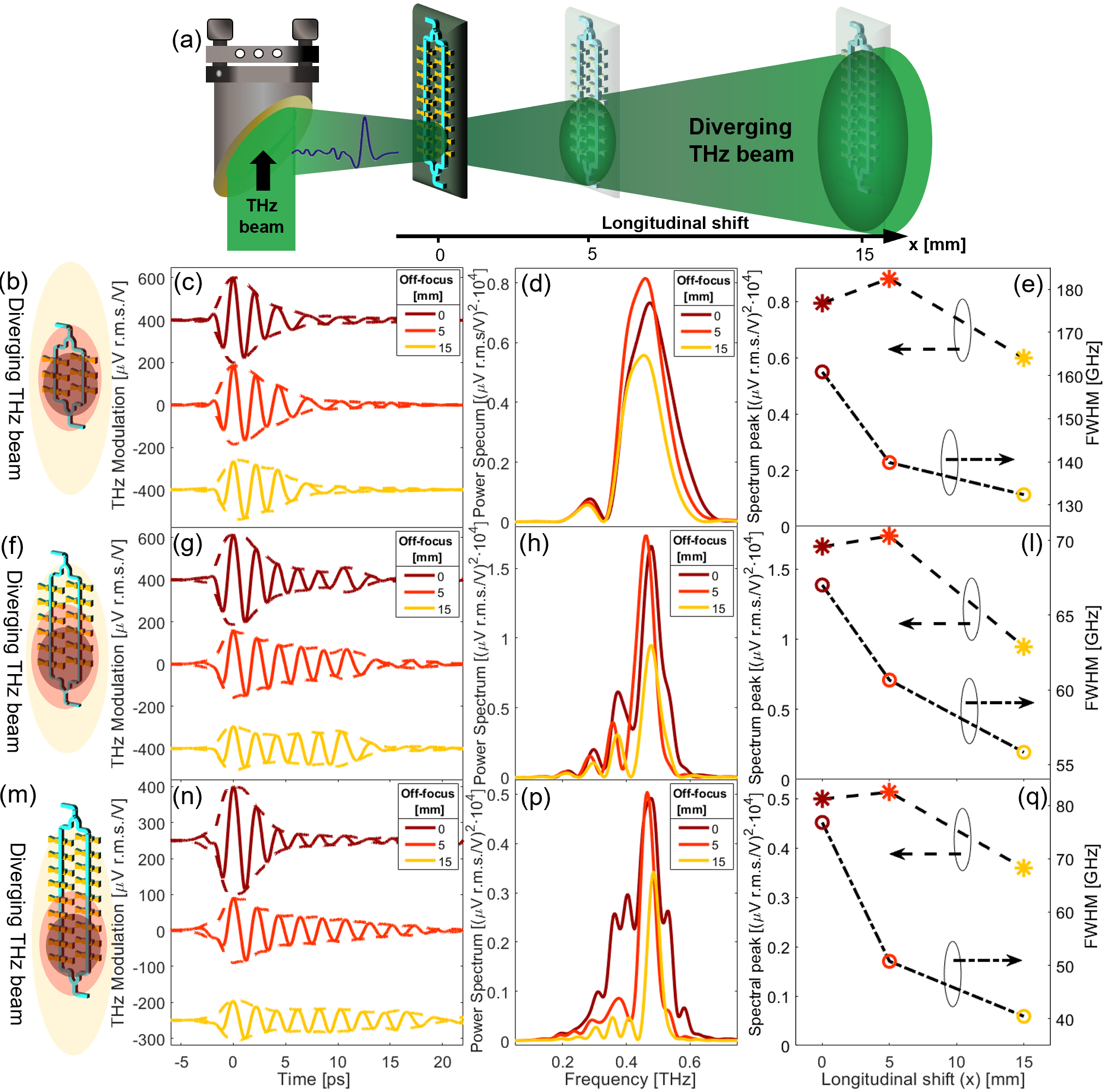}
 \renewcommand{\baselinestretch}{1} 
  \caption{\textbf{Detection under off-centered illumination for increasingly long antenna arrays.} \textbf{(a)} experimental configuration for detecting a diverging THz beam. The device is initially placed on the THz focal plane, where the THz spot mainly illuminates the initial pairs of antennas. The device is then moved out of the focal plane on the z-axis. Time-domain acquisitions have been carried out at positions z = 0, 5, and 15 mm, away from the focal plane. Three different detector types have been used, featuring \textbf{(b)} 3, \textbf{(f)} 6, and \textbf{(m)} 9 antennas per array. Time traces are reported in panel \textbf{(c)-(g)-(n)} for the above-mentioned devices, whereas \textbf{(d)-(h)-(p)} shows the corresponding Fourier-Transform calculated spectra. The general trend among all cases is that the THz transients acquire more cycles as the detectors get farther from the focal plane, due to the broadening THz illumination, which covers an increasing number of antennas. Besides, all spectra reveal that at the intermediate z-position of 5 mm, the peak frequency exhibits a higher amplitude, due to an augmented contribution from more antennas, despite the decrease in electric field strength of the diffracting THz beam, compared to the focal plane. These considerations are summarized in plots \textbf{(e)-(l)-(q)}, which show the trends of the frequency peak and its linewidth as a function of the longitudinal shift. In panels (c)-(g)-(n), curves are vertically shifted for clarity.} 
  \label{fig4}
\end{figure}

\section*{Methods}
\subsection*{Design and fabrication of the MZM devices}\label{Meth:fab}
A graphical representation of the device is presented in Fig. \ref{fig1}1c. The optical circuit is fabricated starting from a 600-nm-thick film of X-cut lithium niobate (LN), bonded onto a ~500-$\mu$m-thick high-resistivity silicon substrate stacked with a 2-$\mu$m-thick buffer layer of thermally-grown silicon dioxide (SiO2). A rib waveguide with a 1.5-$\mu$m-wide core is realized by etching 300 nm of the LN film, with a sidewall angle of $\theta = 63$° \cite{Zhu2021IntegratedNiobate}. The waveguide then splits into two arms through a 50/50 directional coupler so as to form a Mach Zehnder interferometer configuration (MZI). A series of gold bow-tie antennas is deposited across the two arms of the MZI. The two antenna arrays are displaced along their arms by a length dependent upon the operating phase-matching frequency of each device (see main text). The lateral distance between the arms of the MZI is 670 $\mu$m in all devices. The waveguides are separated from the 15/285-nm Ti/Au electrode contacts by a distance of 0.9 $\mu$m-thick, on each side, thus forming a total antenna gap of G = 3.3 $\mu$m. This value has been determined in such a way as to relieve plasmonic losses due to the leaking of the optical mode outside the waveguide core \cite{Herter2023TerahertzPlatform}. The antenna gap is filled with an 800-nm-thick Inductively Coupled Plasma Chemical Vapor Deposition SiO2 layer, which acts as a cladding material for the LN waveguide (see Supplementary Fig.~2). Finally, the optical circuit features a pair of grating couplers realized at each end of the MZI to couple in/out the optical probe beam from/to free space.
\subsection*{THz Time-Domain-Spectroscopy setup}\label{Meth:setup}
The experimental setup is fed by a femtosecond laser oscillator (C-Fiber 780, Menlo Systems), free-space coupled and providing two beamlines, one at 1560 nm (first harmonic) and another one at 780 nm (second harmonic) wavelength. The 780 nm line excites an LT-GaAs PCA antenna, emitting THz pulses while biased by a 12-V-square wave voltage, oscillating at 5 kHz. A series of four 90°, off-axis parabolic mirrors collect, collimate, and refocus the THz beam onto the final detector under nearly diffraction limit conditions. We experimentally retrieved a spot size radius of ~0.7 mm at 0.5 THz (see section 3 of Supplementary Information). The 1560 nm line acts as the probe beam. Because of the operation with the integrated devices, for the sake of a fair comparison between the detection methods, the probe beam is fiber-coupled to a standard 2-m-long single-mode fiber (SMF28). When operating the FS-EOS to acquire the reference THz pulse, the probe beam is again coupled in free space and sent to a <110> 3-mm-thick GaAs crystal, which provides a superior collinear phase-matching condition at the 1560 nm wavelength, compared to, e.g., a ZnTe crystal (see Fig. \ref{fig1}). After interaction with the THz wave, the probe beam is acquired by an amplified, balanced photodiode pair (Nirvana, Newfocus), connected to a lock-in amplifier (UHFI, Zurich Instrument). The latter is synchronized to the PCA bias modulation frequency. When the OC-EOS technique is operated, the GaAs crystal is replaced by the MZI devices, while the 2-m-long fiber carrying the probe beam is terminated on an etched fiber tip, which illuminates the grating coupler. A second etched fiber tip collects the probe beam out-coupled from the second grating placed at the opposite side of the optical circuit sends it to a single channel of the balanced photodetector, which now operates in the single-ended configuration. Lock-in acquisition is carried out at the bias modulation frequency, as for the FS-EOS case. 
The integrated devices were fed with an estimated 4-mW-probe power, coupled through gratings and directly traveling along the waveguide. The THz waveforms were recorded by acquiring the readout signal generated by the photodiode while scanning the temporal delay between the probe and THz pulses. 

\vspace{1cm}

\textbf{Data Availability}
The data generated in this study will be made publicly available in the Zenodo database prior to publication.

\medskip
\textbf{Code Availability}
The code used to plot the data within this paper is available in the Zenodo database prior to publication.

\section*{References}
\bibliography{references_MZM_TFLN}

\medskip
\textbf{Acknowledgements}
A.T. and I.C.B.C. acknowledge funding from the European Union’s Horizon Europe research and innovation programme under project MIRAQLS with grant agreement No 101070700 and from the Swiss National Science Foundation (SPARK, grant number 221119). I.C.B.C. acknowledges funding from the Swiss National Science Foundation - National Science Foundation (SNSF-NSF) Lead Agency program under award number 219409. A.S.-A. and M.L. acknowledge funding from the National Science Foundation - Swiss National Science Foundation (NSF- SNSF) ECCS-2407727 and DARPA LUMOS program HR001120C0137 Defense Advanced Research Projects Agency (HR0011-20-C-0137). The fabrication of these chips was performed in part at the Center for Nanoscale Systems (CNS), a member of the National Nanotechnology Coordinated Infrastructure Network (NNCI), which is supported by the National Science Foundation under NSF Award no. 1541959. 

\medskip
\textbf{Author contributions} I.C.B.C., A.T. and A.S.A conceptualized the project. A.T. built the experimental setup and performed the terahertz measurements. A.S.A. designed and fabricated the samples. A.T. and I.C.B.C performed the CST simulations and analyzed the data. A.T., A.S.A. and I.C.B.C. wrote the manuscript with feedback from all authors. The work was done under the supervision of I.C.B.C and M.L. 

\medskip
\textbf{Competing interests}
M.L. is involved in developing TFLN based technologies at HyperLight Corporation.

\medskip
\textbf{Disclaimer}
The views, opinions and/or findings expressed are those of the author and should not be interpreted as representing the official views or policies of the Department of Defense or the U.S. Government.The authors declare no competing interests. 

\textbf{Corresponding authors} Correspondence to Alessandro Tomasino (alessandro.tomasino@epfl.ch), or Ileana-Cristina Benea-Chelmus (cristina.benea@epfl.ch). 
%\listoffigures

\newpage
\pagenumbering{gobble}
\centering{\textbf{\Large Supplementary information for} \\  Large-area photonic circuits for terahertz detection and beam profiling}\\
\date{}
{\small A. Tomasino, A Shams-Ansari, M. Lončar, I.-C. Benea-Chelmus}

\newpage

\includepdf[pages=-]{supplementary.pdf}

\end{document}

\renewcommand\refname{Supplementary References}

\maketitle
\newpage
\tableofcontents
\newpage
\section{Supplementary Note 1: Device parameters}\label{Supp:Optical_param}
The fabrication process of our samples is explained in the methods section. The dimensions of the samples are listed in the Supplementary Table~\ref{tab:parameter}. For some antenna parameters, we give a certain range, since these values were changed for different investigations.
\begin{table}[htb]\centering
    \centering
    \begin{tabular}{l|c|r}
         thickness of silicon substrate: & \hspace{0.5cm}$H_\mathrm{Si}\hspace{0.5cm}$ & 500\,$\upmu$m\\
         thickness of silicon oxide isolation layer:\hspace{0.5cm} & $H_\mathrm{SiO_2}$ & $4.7\,\upmu$m \\
         thickness of lithium niobate slab layer: & $H_\mathrm{slab}$ & $300\,$nm \\
         thickness of silicon oxide cladding layer: & $H_\mathrm{clad}$ & $1100\,$nm \\
         gold thickness: & $H_\mathrm{Au}$ & 300\,nm\\
         length of bow-tie arm: & $L_\mathrm{ant}$ & \hspace{0.5cm}90\,$\upmu$m \\
         inner bow-tie arm width: & $w_\mathrm{ant}$ & 5\,$\upmu$m\\
         outer bow-tie arm width: & $W_\mathrm{ant}$ & 30\,$\upmu$m\\
         gap length: & $l_\mathrm{gap}$ & 60$\,\upmu$m \\
         electrode gap width: & $w_\mathrm{bar}$ & 2\,$\upmu$m\\
         antenna gap width: & $w_\mathrm{gap}$ & 3\,$\upmu$m\\
         waveguide width: & $w_\mathrm{wg}$ & 1.5\,$\upmu$m\\
         waveguide height: & $H_\mathrm{wg}$ & $600\,$nm \\
         distance between two antennas (same arm): & $D_1$ & $266\,\upmu$m\\
         distance between two antennas (opposite arm): & $D_2$ & $133\,\upmu$m\\
         length of the antenna array: & $L_{MZM}$ & $2.6\, $mm\\
         width of the MZM: & $W_{MZM}$ & $677\, \upmu$m\\
    \end{tabular}
    \caption{List of antenna and waveguide design parameters.}
    \label{tab:parameter}
\end{table}

\begin{figure}[htb]
 \includegraphics[width=17cm]{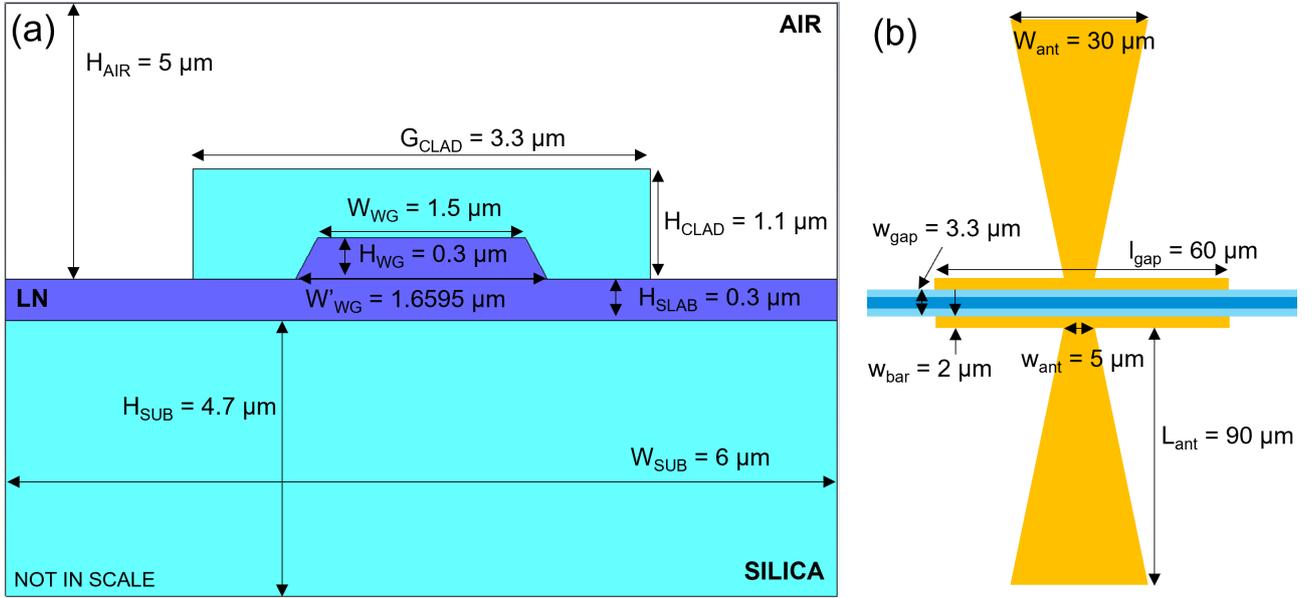} 
 \centering
 \renewcommand{\baselinestretch}{1}
   \caption{Schematics (a) of the waveguide geometry used to perform simulations in CST Microwave Studio and (b) of the top view of the THz antenna layout. All numbers are listed in the table \ref{tab:parameter}.}
  \label{fig1_schematics}
\end{figure}

\begin{figure}[htb]
 \includegraphics[width=17.5cm]{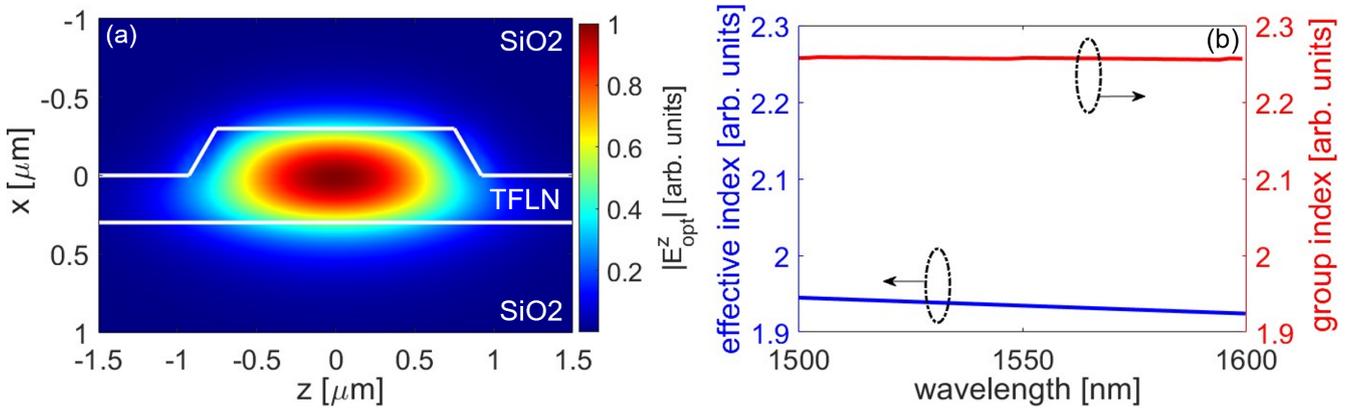} 
 \centering
 \renewcommand{\baselinestretch}{1}
   \caption{(a) Simulated electric field profile of the fundamental mode supported by the TFLN waveguide. The mode is polarized along the z-axis. (b) Effective refractive index (blue solid line) and group index (red solid line) of the fundamental mode as a function of the wavelength in the telecom range. }
  \label{fig2_optical_mode}
\end{figure}

\section{Supplementary Note 2: Terahertz Time Domain Spectroscopy setups for experimental characterizations}\label{Supp-TDS}
A simplified sketch of the two-color time-domain spectroscopy setup is shown in Supplementary Fig. \ref{fig3_setup}. The Menlo C-780 fiber laser provides a train of 60-fs-long optical pulses (with an extremely broad spectrum ranging from 1420 nm to 1600 nm) and a peak average power of 550 mW at a 100 MHz repetition rate. The laser also provides a second output at the second harmonics (780 nm) of the nominal fundamental wavelength (1560 nm). Since they both originate from the main beam, the timing of the two pulses at the two different colors can be easily retrieved. Therefore, we use the 780 nm beamline to pump an LT-GaAs photoconductive (PCA) antenna, which acts as a THz source. The average pump power used is 100 mW, loosely focused to cover the active area of the PCA. The antenna is biased with a bipolar square wave having a peak-to-peak amplitude of 12 V, on-off modulated at 5 kHz. The emitted THz pulsed beam is collected through a 2-inch-diameter off-axis parabolic mirror (OAM1), with a 2-inch focal length. The latter forms a THz beam with a mean diameter of around 1.2 cm. Such a THz beam is subsequently handled by a series of parabolic mirrors (OAM2-3-4), which expand the beam diameter by a factor of two before being tightly focusing it onto the final detector. On the detection side, the probe beam is coupled into a single-mode fiber (SMF28) and then coupled out again in air to carry out free-space electro-optic sampling (FS-EOS) in a 1-mm-long <110> GaAs crystal (indicated as EOX in Fig. \ref{fig3_setup}). The probe beam is focused using a 30-cm-lens and then re-collimated - after interacting with the THz pulse - by a 5-cm-lens, thus ensuring a complete illumination of the sensitive area of the balanced photodiode pair (BPD, Nirvana, Newfocus). Acquisition of the reference THz waveform is performed via lock-in detection (synchronized to the bias modulation frequency, 5 kHz) of the differential signal generated by the BPD while scanning the delay between the THz and the probe pulses. The latter is mechanically introduced by a delay stage placed on the optical pump path and controlled through a software application. \\
To carry out the on-chip electro-optic sampling, the probe beam path is modified as depicted in Supplementary Fig. \ref{fig4}. Specifically, we bypassed the second fiber coupler, and directly connected the SMF fiber to a free-standing “fiber-probe”. The latter consists of a short bare fiber patch that terminates with a cleaved facet. This way, the light beam emitted from the fiber probe shines on top of the grating coupler realized on the chip. To do so, we used a piezoelectric module that enables extremely fine movements with a step size in the order of nanometers. A CCD camera images the chip, allowing the operator to pinpoint the position of the grating couplers, as well as keep track of the relative position between the MZM device and the focal point of the parabolic mirror (which, in turn, indicates the position of the THz beam spot). 

\begin{figure}[htb]
 \includegraphics[width=17.5cm]{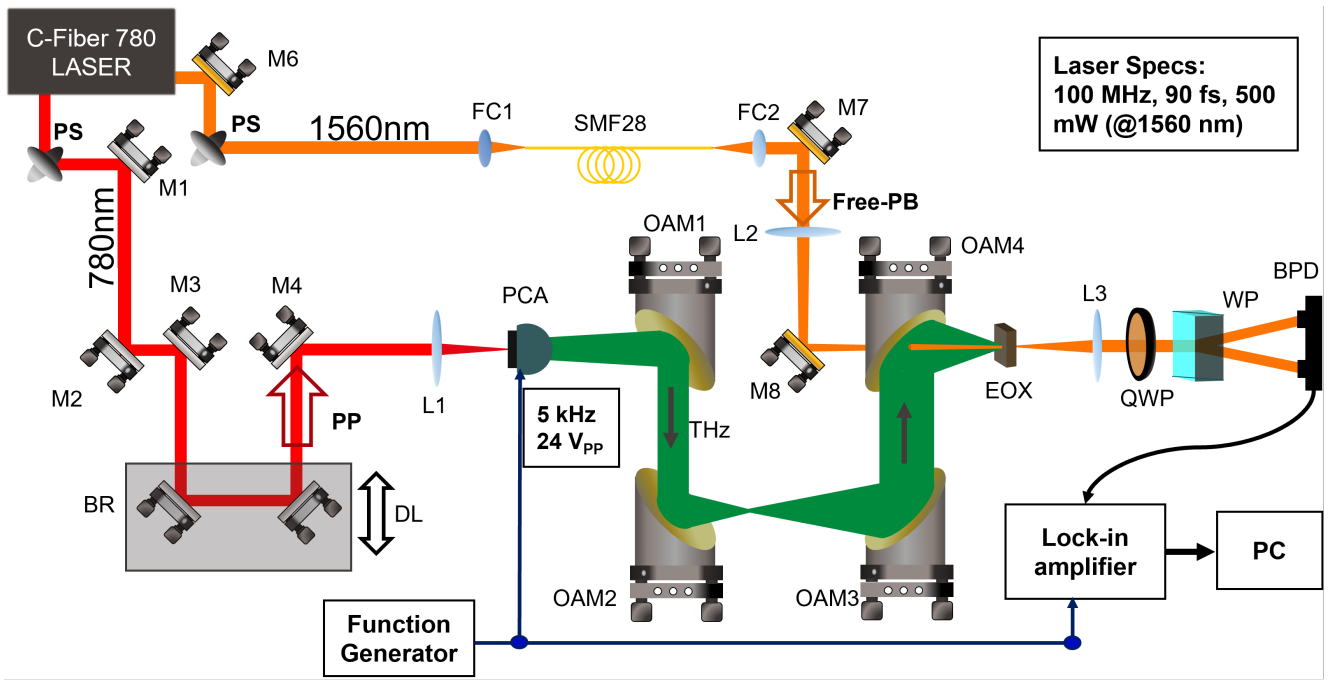} \renewcommand{\baselinestretch}{1}
 \centering
   \caption{\textbf{Two-color THz Time-Domain Spectroscopy set-up.} An optical beam at a wavelength of 780 nm drives a photoconductive antenna (PCA) emitting a burst of THz pulses. The THz beam is handled by a series of 4 parabolic mirrors (OAM) and finally focused onto a bulk electro-optic crystal (EOX). The 1550-nm probe beam overlaps in time and space with the THz beam into the EOX crystal to carry out free-space electro-optic sampling. The THz-induced polarization modulation of the probe beam is revealed by using a quarter waveplate (QWP) and a Wollaston prism (WP), which split and send the two polarization components to a balanced photodiode pair (BPD). A function generation feeds the PCA with a square wave bias voltage, also providing the reference signal for the lock-in amplifier that acquires the readout signal from the BPD.} 
  \label{fig3_setup}
\end{figure}

\begin{figure}[htb]
 \includegraphics[width=17.5cm]{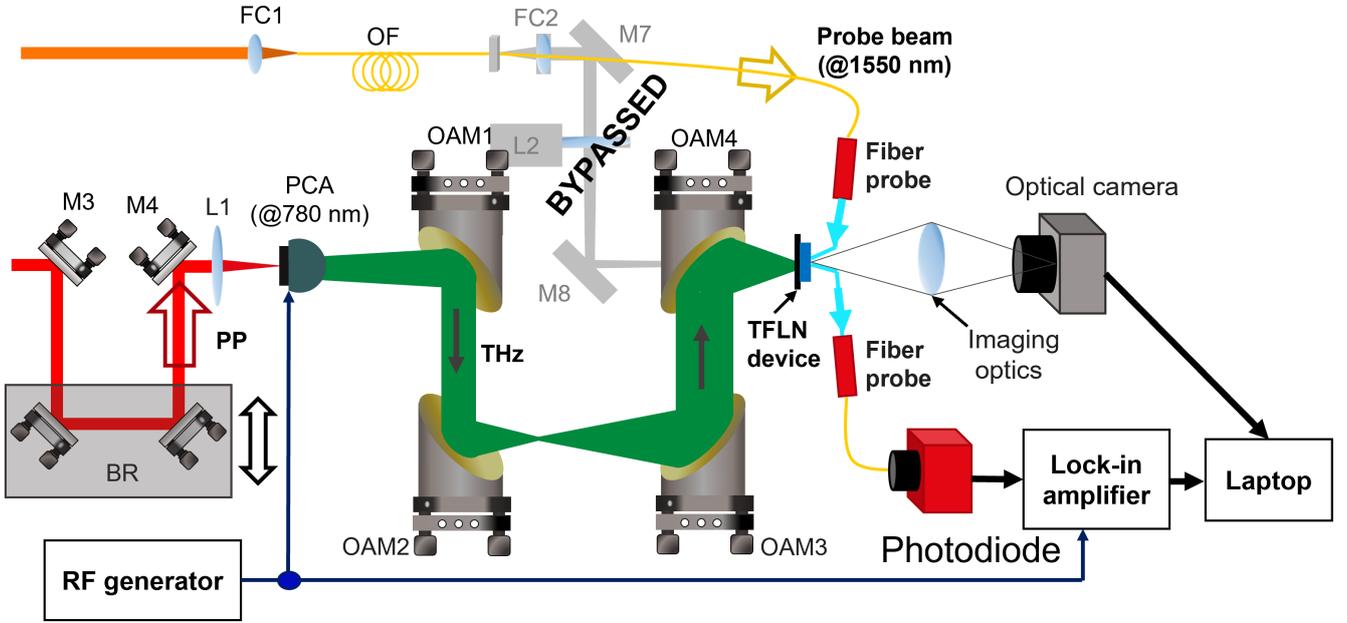} \renewcommand{\baselinestretch}{1}
 \centering
   \caption{\textbf{Modified two-color THz Time-Domain Spectroscopy set-up for on-chip electro-optic sampling.} The generation path is unchanged. On the detection side, the probe beam is coupled in and out of the chip through the fiber probes placed on top of grating couplers. The chip is placed on the THz focus and imaged via a CCD camera to allow alignment of the fiber probes. The out-coupled probe beam is sent to a single-ended photodiode for acquisition. } 
  \label{fig3_setup}
\end{figure}

The out-coupled probe beam is collected through the use of a second fiber probe, controlled by a second piezoelectric stage, and operated reciprocally with respect to the previous one. The light is then sent to a single photodiode. Note that no differential technique could be implemented in this configuration as the output of the interferometer is single-ended. Acquisition of the THz waveform is performed in the same manner as the free-space case. 

\section{Supplementary Note 3: Spatio-temporal knife-edge technique for THz beam profiling}\label{Supp:Knife-edge}
In this section, we explain how we characterized the THz beam in terms of the radial dimension associated with each frequency component. To do so, we carried out a modified version of the well-known knife-edge technique for conventional optical beams \cite{Peccianti2013ExactMeasurements}. We recall that a traditional knife-edge technique \cite{DeAraujo2009MeasurementAnalysis} assumes that the emitted beam exhibits an electric field profile with separable temporal and spatial distributions. This is usually the case for a beam either emitted by a large-area source (i.e., larger than the wavelength squared) or collimated with very long Rayleigh ranges. In this case, the intensity profile can be decomposed into the product of two functions solely dependent upon one single transverse coordinate. In formulae, if the beam propagates along the z direction and its intensity profile $I(x,y)$ is distributed on the \textit{xy}-plane, the separability condition allows to write: $I(x,y) = I_x(x)I_y(y)$. When a blade is inserted in the \textit{xy}-plane and cuts the beam, for instance, along the x-axis up to the coordinate $x_0$, the power $P(x_0)$ unblocked by the blade and reaching an optical detector can be written as:
\begin{align}
    P(x_0) = \int_{-\infty}^{x_0}I_x(x)dx\int_{-\infty}^{+\infty}I_y(y)dy
    \label{energy_x0}
\end{align}
By varying $x_0$ through the translation of the blade across the entire beam size, the function $P(x_0)$ can be reconstructed (with $x_0$ being now a continuum variable). The intensity profile can be then retrieved by differentiation of Eq. \ref{energy_x0}. Using the same procedure along the y-axis, it is possible to reconstruct $P(y)$ and the associated intensity profile, thus providing the entire characterization of the beam dimensions. The application of this technique to THz beams is generally not of an immediate implementation. Indeed, owing to the sub-wavelength geometries often adopted to generate THz beams, the standard knife-edge technique fails to reproduce the exact spatial profile, since the temporal and spatial coordinates couples along the propagation. However, in this work, the THz beam is emitted by a large aperture PCA provided with a hyper-hemispherical silicon lens that forms an emerging beam with a size comparable to the lens radius (around 0.5 cm). In addition, the series of parabolic mirrors in our setup forms a THz beam with a diameter of a few centimeters, thus making the use of the knife-edge technique possible. Furthermore, the access to the electric field waveforms through TDS measurements (i.e., to both its amplitude and phase) can be exploited to perform a superior implementation compared to the optical case. More in detail, in the field-resolved system depicted in Fig.\ref{fig3_setup} of the main manuscript, where the transverse plane is \textit{zy}, when the blade cuts a spatio-temporal THz beam, the uncut section of the beam becomes a new source of radiation with a resultant electric field $E^{res}_{THz}(z_0,t)$ proportional to the incident electric field $E^{in}_{THz}(z,t)$ as:
\begin{align}
    E^{res}_{THz}(z_0,t) \propto \int_{z_0}^{\infty}E^{in}_{THz}(z,t)dx \propto  \frac{E_0}{2}\textrm{erfc}(\frac{z-z_0}{R_z})
    \label{field_cut}
\end{align}
where $z_0$ is the blade position, while we are assuming that the input THz beam resembles a Gaussian spatial profile ($E^{in}_{THz}(z) \propto E_0\exp[(-z/R_z)^2]$), with $R_z$ being the radius of the collimated beam that we want to estimates.  

\begin{figure}[htb]
 \includegraphics[width=17.5cm]{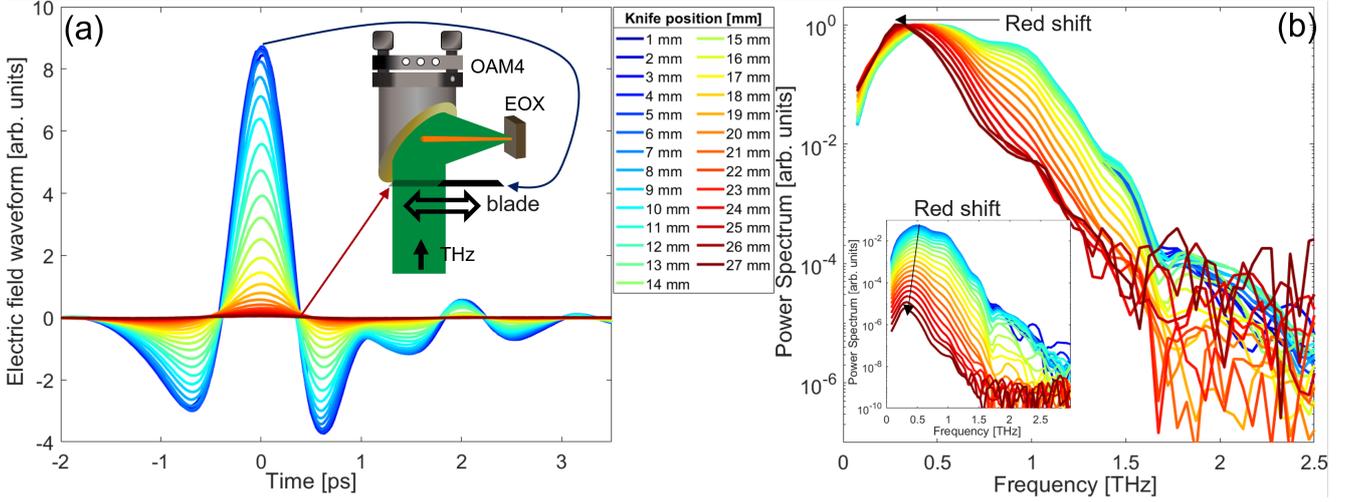} 
 \centering
 \renewcommand{\baselinestretch}{1}
 \centering
   \caption{\textbf{Time-Domain Spectroscopy of a transversely cut THz pulsed beam.} (a) THz waveforms retrieved as a function of the relative position of the blade inside the THz beam. (b) FFT spectra of the waveforms in (a). Curves are normalized with respect to their maximum. The inset shows the non-normalized spectra. As the blade cuts deeper into the beam, the peak of the reconstructed spectra moves towards lower frequencies.}
  \label{Time_blade}
\end{figure}

\begin{figure}[htb]
 \includegraphics[width=17.5cm]{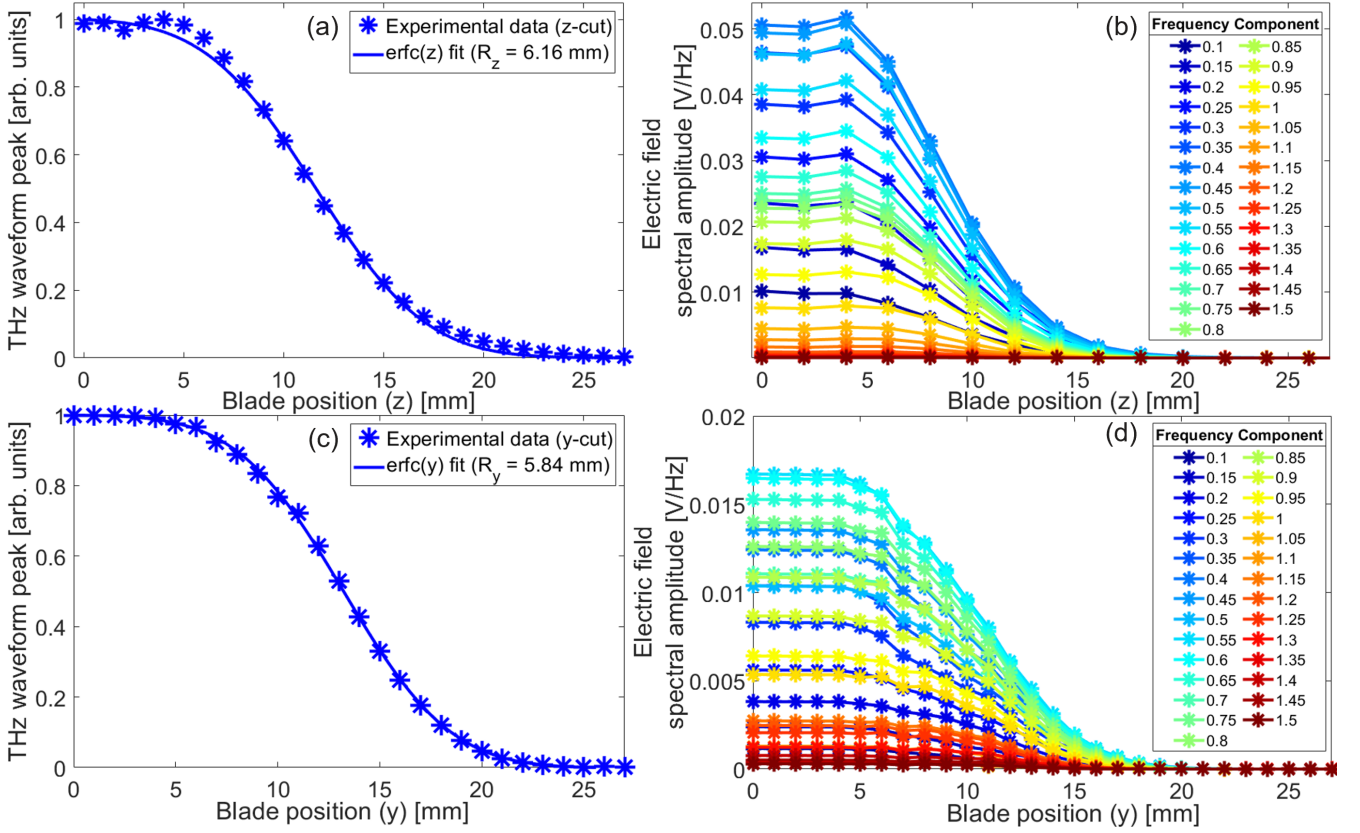} \renewcommand{\baselinestretch}{1}
 \centering
   \caption{\textbf{Knife-edge characterization of the THz beam.} Amplitude peak of the THz waveform recorded as a function of the blade position (blue dots) fitted with the theoretical error function ('erfc') characterizing ideal Gaussian beams (blue solid line) for the case of horizontal cut along the z-axis (a) and vertical cut along the y-axis (c). (b) Peak value trends of the spectral amplitude for selected frequency components as a function of the blade position, for the (b) horizontal and (d) vertical cut.} 
  \label{Fit_space_and_freq}
\end{figure}

\begin{figure}[htb]
 \includegraphics[width=17.5cm]{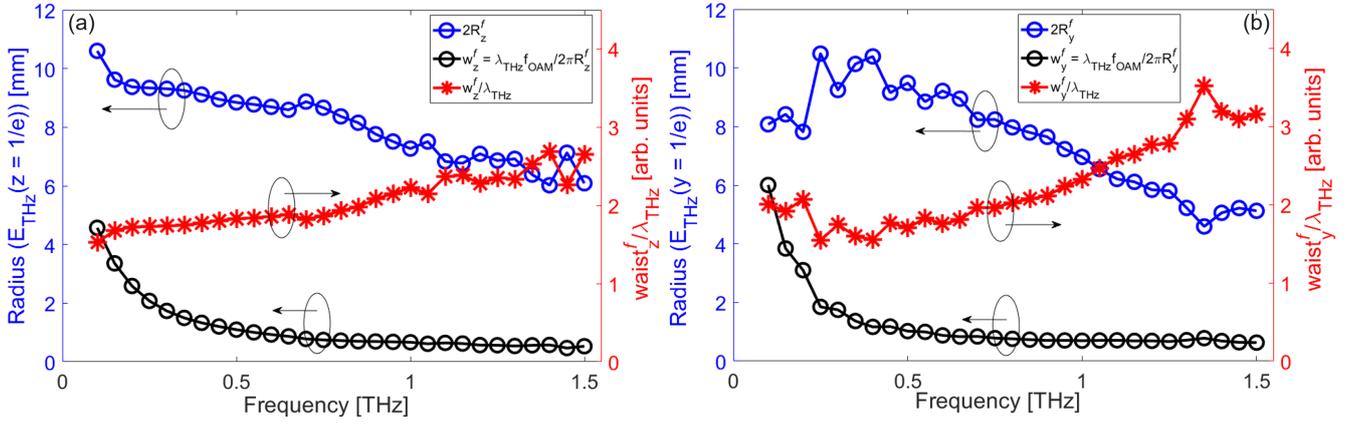} \renewcommand{\baselinestretch}{1}
 \centering
   \caption{\textbf{Reconstruction of the THz beam radii and waist sizes as a function of the frequency.} (a) Calculated collimated beam size (blue dotted line) and waist size (black dotted line) as a function of the frequency. The red stars curve represents the Rayleigh ratio indicating that the THz source operates above the diffraction limit.} 
  \label{Final_charact_KE}
\end{figure}

In Eq. \ref{field_cut}, 'erfc' stands for the complementary error function. The same considerations apply to the \textit{y}-direction, leading to the estimation of $R_y$. Therefore, by performing several acquisitions as a function of the blade position $z_0$, the transverse profile of the collimated THz beam can be retrieved by differentiating Eq. \ref{field_cut} with respect to the \textit{z}-coordinate. In practice, we utilized the FS-EOS configuration to record a series of the THz waveforms for different blade positions inserted in the THz beam path right in front of OAM2, with focal length $f_{OAM}$. This way, the uncut section of the THz beam will be straightforwardly captured and focused by the parabolic mirror into the detection crystal. Supplementary Fig. \ref{Time_blade}a shows the waveforms recorded as a function of blade position, whereas Fig. \ref{Time_blade}b represents the FFT-calculated spectra corresponding to the waveforms in panel (a). Larger numbers for the blade position correspond to a deeper insertion into the THz beam path. Since a THz beam is quite broadband, it could be regarded as a superposition of a virtually infinite number of beams at different frequency components. As lower frequencies diffract out in larger radiation patterns compared to higher frequencies, it is plausible to assume that the periphery of the THz beam is more low-frequency rich compared to its central part. Conversely, the energy associated with low-frequency components is spread across a much larger area. Because of this, as the blade moves deeper into the THz beam, not only does the amplitude of the acquired THz transient diminish (see Supplementary Fig. \ref{Time_blade}a), but also their waveforms and consequently their spectra experience a significant reshaping. In particular, we observe a noticeable red-shift of the calculated peak frequency (Supplementary Fig. \ref{Time_blade}b), which has to be ascribed to the fact that the blade is more efficient in filtering high-frequency components for deeper cuts, because of the reduced spatial extension compared to those at lower frequencies. By taking the peak values of each temporal amplitude and plotting them as a function of the blade position along \textit{z}, we constructed the curve displayed in Supplementary Fig. \ref{Fit_space_and_freq}a. The data points are then fitted with the first integral of a Gaussian function, as described in Eq. \ref{field_cut}. The good fit between the model and data points proves the good quality of the formed THz beam. This fit returns a mean beam radius of $R_z$ = 6.16 mm. The same procedure returns $R_y$ = 5.84 mm. Here, the term ‘mean’ is used to stress the fact that it accounts for an average value across the whole THz spectrum. 
To retrieve more specific values, we take advantage of the time-resolved knife-edge technique. Indeed, similarly to the time data, we can build a graph similar to that in Supplementary Fig. \ref{Fit_space_and_freq}a, yet in a dense range of frequency components, as reported in Supplementary Fig. \ref{Fit_space_and_freq}b. We repeated the same procedure by cutting the beam along the \textit{y}-axis and obtaining the plots in Supplementary Fig. \ref{Fit_space_and_freq}c-d. By fitting the spectral curves in Supplementary Fig. \ref{Fit_space_and_freq}b-d with different 'erfc' functions, we can reconstruct a trend of the beam radius as a function of the THz frequency components, which are reported in Supplementary Fig. \ref{Final_charact_KE}a and b, for the \textit{z}-axis and \textit{y}-axis, respectively. It results that the THz beam has a frequency-dependent beam radius ($R^{f}_z$) slowly decreasing with the frequency, ranging approximately between $R^{f}_z$ = 4.5 mm - 8 mm, in the spectral window between 0.2 and 1.5 THz. Finally, by using the laws of Fraunhofer diffraction, we can reconstruct the profile of the THz electric field at the focal point of the parabolic mirror, by taking the spatial Fourier transform of the collimated beam profile (i.e., the spatial derivative of Eq. \ref{field_cut}):
\begin{align}
    E^{meas}_{THz}(z',t) = \int_{-\infty}^{+\infty}E^{in}_{THz}(z,t)e^{-ik_zz}dz \propto \int_{-\infty}^{+\infty} E_0e^{(-z/R_z)^2}e^{-ik_zz}dz \propto \exp{\left[\left(-\frac{\pi R_z z'}{f_{OAM}\lambda}\right)^2\right]}
    \label{field_cut_FS}
\end{align}
where $k_z = kz'/f_{OAM}$, being $k = 2\pi /\lambda$ the THz wavevector, and $z'$ the conjugate spatial coordinate of \textit{z} in the focal plane of the parabolic mirror (reciprocal plane). From Eq. \ref{field_cut_FS}, we derive that the mean waist size $w^{z}_{THz}$ at the final focal plane can be calculated as: $w^{z}_{THz} = \frac{\lambda f_{OAM}}{\pi R_z} = 660 \mu m $. Once again, a similar equation to that in Eq. \ref{field_cut_FS} can be utilized for each frequency component, which allows to reconstruct the trend of the waist size as a function of the frequency. The latter is reported in Supplementary Fig. \ref{Final_charact_KE}a, showing a THz waist ranging between 3.2 - 0.375 mm in the 0.2 -1.5 THz range along the z-axis. Supplementary Fig. \ref{Final_charact_KE}b shows the frequency dependent waist size calculated for the for the y-axis, showing slightly larger values than the z-axis case, due to the fact that the collimated beam radius are in opposite relationship (i.e., $R_y < R_z$). The mean waist size along the y direction is $w^{y}_{THz} = \frac{\lambda f_{OAM}}{\pi R_y} = 700 \mu m $. These two values are reported in Fig. \ref{fig3}h of the main manuscript.

\section{Supplementary Note 4: Analytical modeling of the quasi-phase-matching mechanism occurring in arrays of THz antennas}\label{Supp:MZM_theory}
In this section, we derive the spectral response of the TFLN device in terms of its intensity transmission function $T_{MZM}$ reported in Eq.\ref{array_freq_resp} of the main manuscript. We start the digression by considering the scheme in Supplementary Fig. \ref{MZM_config}a. It depicts a conventional MZM geometry, where each arm imparts a different phase retardation to each of the two optical beams guided through two distinct components, represented as the red (up) and green (low) blocks. Here, we assume that the input optical beam has an electric field $E_0^{in}$, which splits into two components with identical initial amplitude $E_0^{in}/\sqrt{2}$ and phase (here, conveniently fixed at zero for simplicity). After propagating along each arm, the THz electric field established in the antenna arrays will modulate the two optical beams as:
\begin{align}
    E^{U} = \frac{E_0^{in}}{\sqrt{2}}e^{i(\phi_{THz}^{U}+\phi_{Q})}\\
    E^{D} = \frac{E_0^{in}}{\sqrt{2}}e^{i\phi_{THz}^{D}}
\end{align}
where $\phi_{THz}^{U}$ and $\phi_{THz}^{D}$ are the phase retardation experienced by the upper and lower beam, respectively. Here, we are still treating a generic case where the two modulations might differ between arms. The term $\phi_{Q}$ accounts for the built-in dephasing between the two arms, intentionally introduced to operate the MZM device at its quadrature point (i.e., with its output intensity equal to half of that of the input). At the MZM output, the superposition of the two optical beams will give rise to:
\begin{align}
    E_{out} = E^{U}+E^{D} = \frac{E_0^{in}}{2}\left[e^{i(\phi_{THz}^{U}+\phi_{Q})}+e^{i\phi_{THz}^{D}}\right] = \frac{E_0^{in}}{2}e^{i\frac{\phi_{THz}^{U}+\phi_{Q}+\phi_{THz}^{D}}{2}}\left[e^{i\frac{\phi_{THz}^{U}+\phi_{Q}-\phi_{THz}^D}{2}}+e^{-i\frac{\phi_{THz}^{U}+\phi_{Q}-\phi_{THz}^{D}}{2}}\right]
\end{align}
The quadrature point correspond to the condition $\phi_{Q} = \pi/2$, thus, after applying further simplifications, we achieve:
\begin{align}
    E_{out} = E_0^{in}e^{\frac{i\pi}{4}}e^{i\frac{\phi_{THz}^{U}+\phi_{THz}^{D}}{2}}\cos{\left[\frac{\phi_{THz}^{U}-\phi_{THz}^{D}}{2}+\frac{\pi}{4}\right]}
    \label{inteference_electric}
\end{align}
Since the photodetector is sensitive to the intensity of the probe electric field in Eq. \ref{inteference_electric}, we can estimate it as:
\begin{align}
    I_{out} \propto |E_{out}|^2 = |E_0^{in}|^2\cos^2{\left[\frac{\phi_{THz}^{U}-\phi_{THz}^{D}}{2}+\frac{\pi}{4}\right]} = \frac{|E_0^{in}|^2}{2}\left[1+\cos{\left(\phi_{THz}^{U}-\phi_{THz}^{D}+\frac{\pi}{2}\right)}\right]
\end{align}
Now, by defining $\Delta \phi = \phi_{THz}^{U}-\phi_{THz}^{D}$ and using $I_{in} \propto |E_0^{in}|^2$, we can finally write:
\begin{align}
    I_{out} = \frac{I_{0}^{in}}{2}\left[1+\sin{\Delta \phi}\right]
\end{align}

\begin{figure}[htb]
 \includegraphics[width=17.5cm]{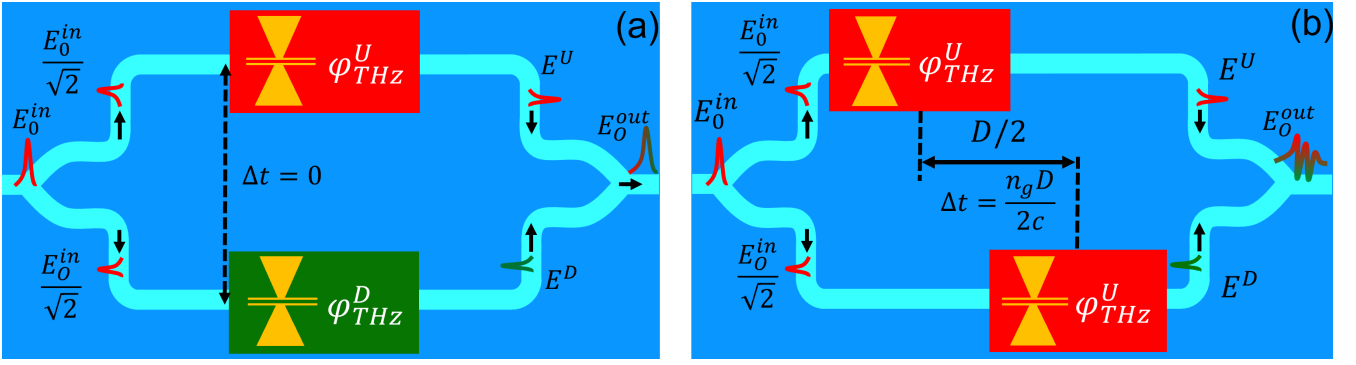} \renewcommand{\baselinestretch}{1}
 \centering
   \caption{\textbf{Electro-optic modulation in a Mach Zehnder interferometer hosting arrays of THz antennas.} The input optical beam  $E_0^{in}$ is split into two sub-beams with identical initial amplitude and phase. (a) In the symmetric case, the two sub-beams undergo the same phase modulation ($\phi^{U}_{THz}$ and $\phi^{D}_{THz}$ for the upper and lower arm, respectively) as each beam encounter the array on (red and green boxes) at the same time instant. The two sub-beam recombine generating an output beam that carries twice the phase modulation of the single arm case. (b) In the displaced case, the same array is translated along the arm by a length D/2, so that the lower sub-beam has to travel for an extra time delay $\Delta t $ before interacting with the array. The unbalancing of the two different phase modulations leads to the amplitude modulation of the recombined beam at the output.} 
  \label{MZM_config}
\end{figure}

By taking the difference between the readout intensity measured with and without THz-induced modulation $\Delta \phi$, we can write the THz-induced intensity modulation as:
\begin{align}
    \Delta I_{out} =  I_{out}\left(E_{THz} \neq 0\right) - I_{out}\left(E_{THz} = 0\right) =  \frac{I_{0}^{in}}{2}\sin{\Delta \phi} \approx \frac{I_{0}^{in}}{2}\Delta \phi 
    \label{modulation}
\end{align} 
where we have used the fact that $\Delta \phi \ll 1$ under weak THz modulation.

\begin{figure}[htb]
 \includegraphics[width=15cm]{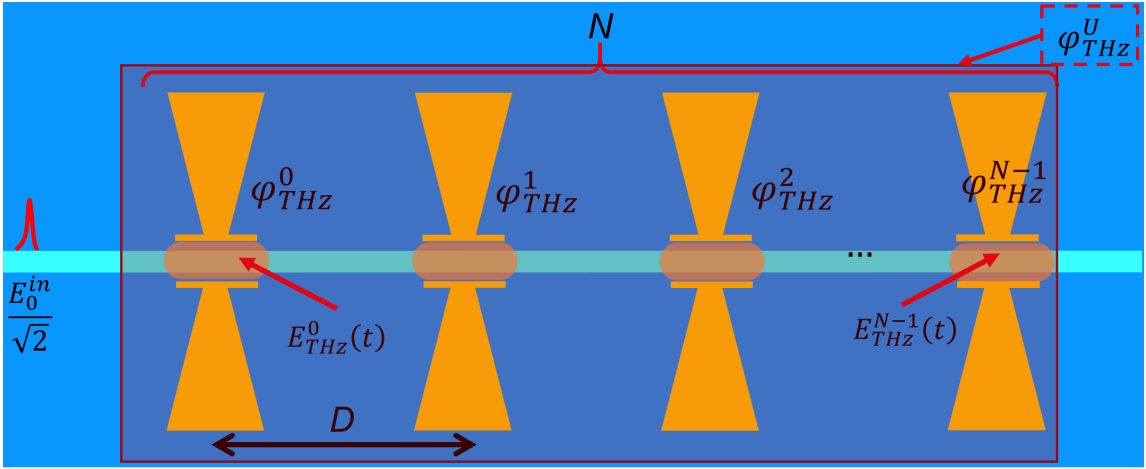} 
 \centering
 \renewcommand{\baselinestretch}{1}
   \caption{\textbf{Building-up of the phase modulation in an array of THz antennas.} The sub-probe propagates through an optical waveguide that crosses the gaps of an array of \textit{N} antennas with a spatial period \textit{D} (semi-transparent red box). At each antenna encounter, the probe beam experiences a phase modulation $\phi^n_{THz}, \text{with} n = 0,...,N-1$ imparted by the instantaneous THz electric field $E^n_{THz}(t), \text{with} n = 0,...,N-1$ established at each gap. The summation over all contribution construct the total $\phi^U_{THz}$ due to a single arm of the interferometer.} 
  \label{single_array}
\end{figure}

Equation \ref{modulation} states the linear dependence of the probe intensity variation upon the THz electric field, implying a coherent reconstruction of the THz wave. Let us now consider the case that the phase retardation is imparted to the two probe beams at a different time instant, as indicated in Supplementary Fig. \ref{MZM_config}b. In particular, the two blocks are spaced by a distance $D/2$, which corresponds to a time delay of $\frac{T}{2} = \frac{n_gD}{2c}$, where $n_g$ is the group refractive index, and \textit{c} the speed of light in vacuum. Let us also assume that the two blocks apply the same phase modulation when taken singularly. Then, the lower phase modulation will only differ from the upper arm by an extra time delay, namely:
\begin{align}
   \phi_{THz}^{D}(t) = \phi_{THz}^U\left(t+\frac{T}{2}\right) 
\end{align}
that can be expressed in the frequency domain as:
\begin{align}
   \phi_{THz}^{D}(\omega_{THz}) = \phi_{THz}^U(\omega_{THz})e^{i\frac{\omega_{THz}T}{2}} 
\end{align}
Therefore, the total phase modulation imparted at the output of the interferometer can be written in the frequency domain as:
\begin{equation}
\begin{split}
   \Delta \phi(\omega_{THz}) = \ &\phi_{THz}^{U}(\omega_{THz}) - \phi_{THz}^U(\omega_{THz})e^{-i\frac{\omega_{THz}T}{2}} =  \phi_{THz}^{U}(\omega_{THz})\left(1-e^{i\frac{\omega_{THz}T}{2}}\right) =  \\ 
   & = \phi_{THz}^{U}(\omega_{THz})\left[-2ie^{i\frac{\omega_{THz}T}{4}}\sin{\frac{\omega_{THz}T}{4}}\right]
\end{split}
   \label{delta_phi_delayed}
\end{equation}

\begin{figure}[htb]
 \includegraphics[width=17.5cm]{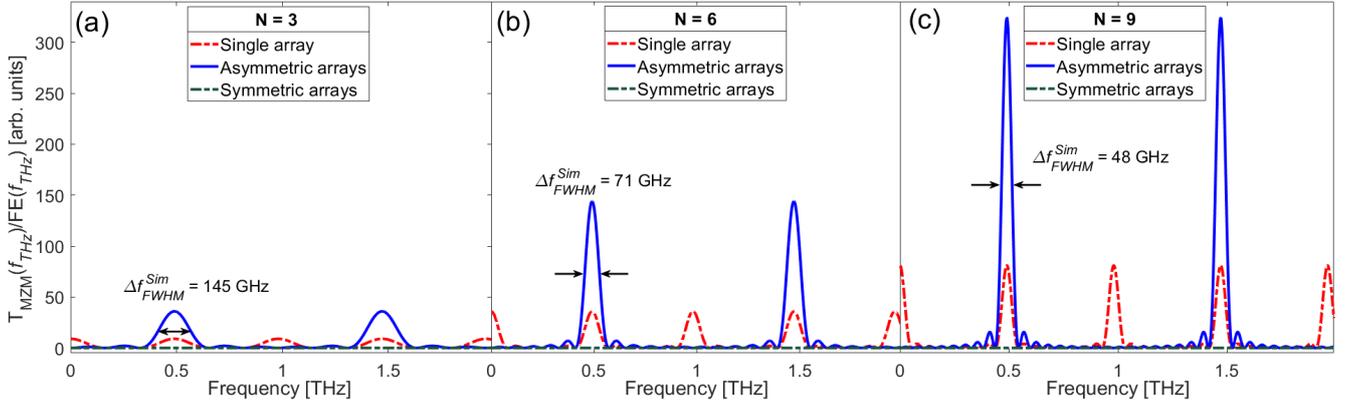} \renewcommand{\baselinestretch}{1}
 \centering
   \caption{\textbf{Spectral transfer function ($T_{MZI}$) of the MZM devices as a function of the number of antennas} red dotted line), double symmetric (green dotted lines) and displaced array (blue solid line) for (c) $N$ = 3, (d) $N$ = 6, (e) $N$ = 9 antennas, as shown in (a) Amplitude peak of the THz waveform as a function of the blade position (blue dots) fitted with the theoretical error function characterizing ideal Gaussian beams (blue solid line). (b) Trend of peak values of selected frequency components as a function of the blade position. (c) Calculated collimated beam size (blue dotted line) and waist size (black dotted line) as a function of the frequency. The red stars curve represents the Rayleigh ratio indicating that the THz source operates above the diffraction limit.} 
  \label{MZM_response}
\end{figure}

Equation \ref{delta_phi_delayed} allows for the computation of the phase retardation due to both arms having the same type of antenna array, at any value of the delay T/2. We note that for $T/2 = 0$ (corresponding to the configuration in Supplementary Fig. \ref{MZM_config}a), the phase imbalance between the two arms is zero ($\Delta \phi(\omega_{THz})$ = 0) at any temporal instant, thus resulting in an identically null intensity modulation of the probe beam, as correctly predicted by Eq. \ref{modulation}. At this point, the only quantity left to be computed is the phase modulation due to a single antenna array. To this end, Supplementary Fig. \ref{single_array} represents the upper sub-probe beam traveling consecutively through the gaps of a \textit{N}-antenna array, separated by a distance equal to \textit{D} from each other. The probe beam crosses each antenna at time instants spaced by the quantity $\Delta t = \frac{n_gD}{c} = T$, so that the delay accumulated increases with steps of $t_n = t_0+nT$, with $t_0$ being the time at which the probe encounters the first antenna and $n = 0, 1, ..., N-1$ is an integer. As shown above, we stress that the delay $T$ is purposely chosen as the double of the delay mutually separating the two opposite arms (i.e., $T/2$). In the time domain, the total phase retardation accumulated by the probe beam at the end of the array is then equal to a summation over all individual contributions: 
\begin{equation}
\begin{split}
\phi_{THz}^{U}(t) = \phi_{THz}(t_0)+\phi_{THz}(t_1)+...+\phi_{THz}(t_{N-1}) = \sum_{n=0}^{N-1} \phi_{THz}(t_0+nT) 
\label{phase_acc_time}
\end{split}
\end{equation}
Moving to the frequency domain, we obtain:
\begin{equation}
\begin{split}
\phi_{THz}^{U}(\omega_{THz}) = \sum_{n=0}^{N-1}\phi_{THz}^{0}(\omega_{THz})e^{in\omega_{THz}T} 
\label{phase_acc_freq}
\end{split}
\end{equation}
where $\phi_{THz}^{0}(\omega_{THz})$ is the Fourier Transform of $\phi_{THz}(t_0)$. We recognize in Eq. \ref{phase_acc_freq} the definition of array factor typical of a phased array \cite{ConstantineA.Balanis2016AntennaEdition}. The sum in Eq. \ref{phase_acc_freq} is simply a geometric series of ratio $e^{i\omega_{THz}T}$ and can be easily calculated as:
\begin{equation}
\begin{split}
\phi_{THz}^{U}(\omega_{THz}) = \phi_{THz}^{0}(\omega_{THz})\frac{e^{i\omega_{THz}NT} - 1}{e^{in\omega_{THz}T} - 1} =  \phi_{THz}^{0}(\omega_{THz})\frac{\sin{\frac{\omega_{THz}NT}{2}}}{\sin{\frac{\omega_{THz}T}{2}}}e^{i\frac{\omega_{THz}T(N-1)}{2}} 
\label{phase_acc_freq_complete}
\end{split}
\end{equation}
By merging the results of Eq. \ref{phase_acc_freq_complete} and Eq. \ref{delta_phi_delayed}, we can write the total phase retardation imparted by interferometer to the output probe beam under THz illumination as:
\begin{equation}
\begin{split}
 \Delta \phi(\omega_{THz}) = -2i\phi_{THz}^{0}(\omega_{THz})e^{-i\frac{\omega_{THz}T}{4}}\frac{\sin{\frac{\omega_{THz}NT}{2}}}{\sin{\frac{\omega_{THz}T}{2}}}\sin{\frac{\omega_{THz}T}{4}}e^{i\frac{\omega_{THz}T(N-1)}{2}} 
\label{phase_difference}
\end{split}
\end{equation}
The information about the THz electric field is included in the term $\phi_{THz}^{0} \propto E_{THz}^{0}$, which also accounts for the spectral response of each antenna through its field enhancement factor $FE(\omega_{THz}) = E^{0}_{THz}/E^{in}_{THz}$, where $E^{in}_{THz}$ is the impinging THz electric field from the free-space. Therefore, we can finally define the transfer function of the MZM with double displaced arrays, normalized to the THz electric field enhanced in each antenna as (Eq. \ref{array_freq_resp_antisymmetric} of the main manuscript):
\begin{equation}
\begin{split}
 \frac{\Delta \phi(\omega_{THz})}{FE(\omega_{THz})} = T_{MZI}(\omega_{THz}) \propto -2ie^{-i\frac{\omega_{THz}T}{4}}\frac{\sin{\frac{\omega_{THz}NT}{2}}}{\sin{\frac{\omega_{THz}T}{2}}}\sin{\frac{\omega_{THz}T}{4}}e^{i\frac{\omega_{THz}T(N-1)}{2}} 
\label{transfer_function}
\end{split}
\end{equation}
Equation.~\ref{transfer_function} can be used to predict the spectral sensitivity of different MZI devices featuring arrays with increasing antennas. More in detail, we here compare the symmetric and asymmetric (i.e., displaced) double array configurations. Supplementary Fig.~\ref{MZM_response} shows the calculated spectral responses for $D_1 = 2D_2 =  266 ~\mathrm{\mu m}$, corresponding to a $f_{PM} = 487 ~\mathrm{GHz}$, and for three different values of $N = 3, 6, 9$, respectively, as shown in Fig. \ref{fig4} of the main text. As expected, the symmetric double array ($\Delta t_{2} = 0$) exhibits no sensitivity over the entire band, regardless of the number of antennas, since $T_{MZI}(f, T/2 = \Delta t_{2} = 0) = 0$. Indeed, in the latter case, the two arms operate identically, effectively acting as a single entity (phase modulator). Therefore, the THz electric field modulation remains encoded into the probe phase, which is not detectable with an optical photodetector. Conversely, the built-in asymmetry of the displaced double array makes the device sensitive to relatively narrow spectral regions around the main $f_{PM}$ and its odd harmonics only. For comparison, the single array case (which also corresponds to an inherent asymmetry between the arms of the interferometer) is displayed in Supplementary Fig.~\ref{MZM_response}. The latter reveals an amplitude response four times smaller (in power) than that of the asymmetric double array, yet with extended sensitivity to the even harmonics. This has to be expected since the $T/2$ displacement featuring the asymmetric configuration leads the even harmonics of the single array case to destructively interfere upon recombination. Finally, we note that in both cases, the linewidth around the frequency components at a multiple of $f_{PM}$ scales inversely with the number of antennas $\Delta f_{FWHM}^{Sim}\propto \frac{1}{N^2}$. This implies that a certain trade-off between detected bandwidth and amplitude response can be achieved by choosing a proper number of antennas. We point out the excellent agreement between the calculated linewidth and that retrieved experimentally, as shown in Fig. \ref{fig2}g of the main manuscript. 

\section{Supplementary Note 5: Temporal interpretation of the formation of the THz transients recorded via antenna arrays.}\label{Supp:Time_formation}
\begin{figure}[htb]
 \includegraphics[width=17cm]{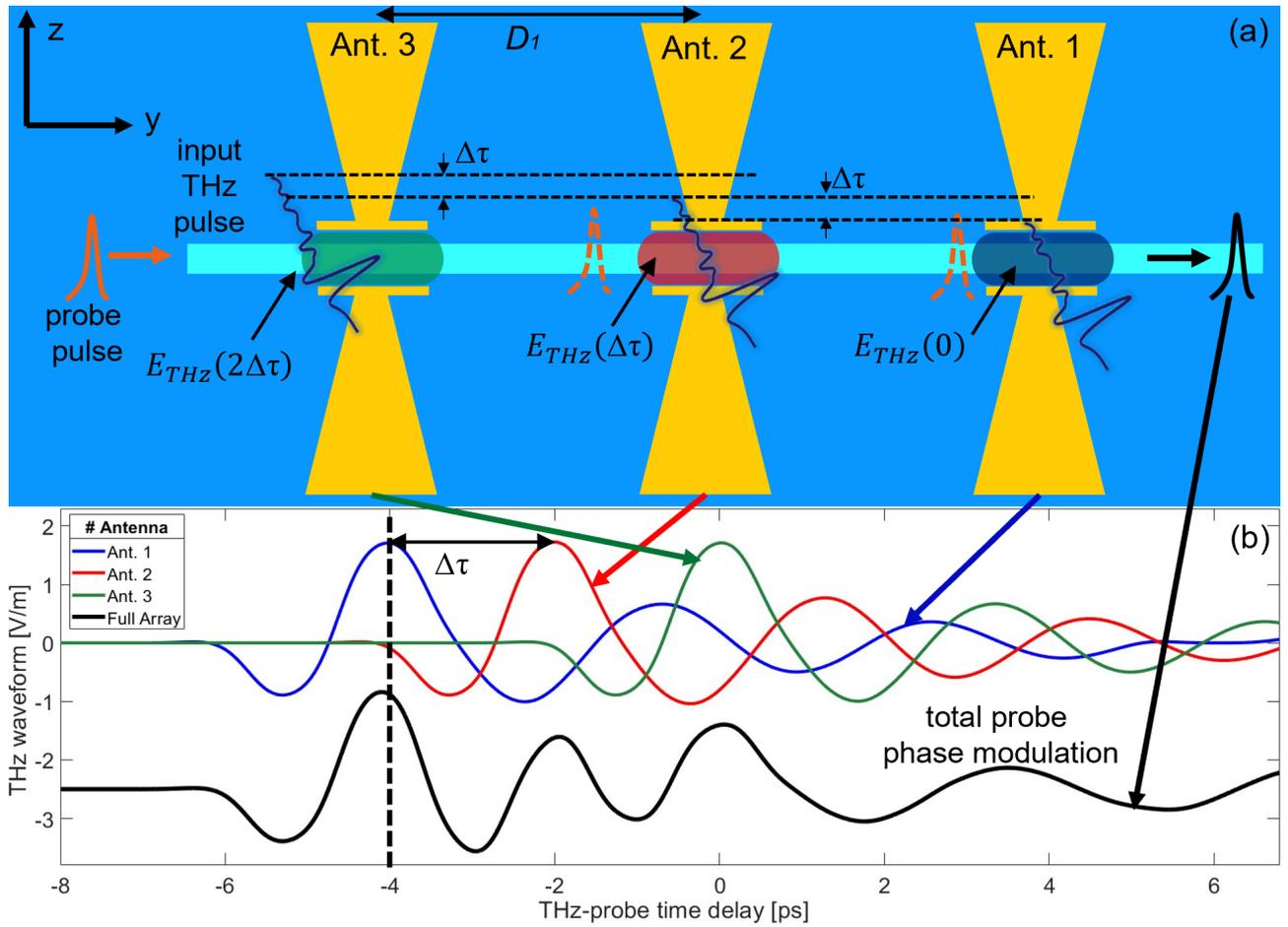} \renewcommand{\baselinestretch}{1}
 \centering
   \caption{\textbf{Time diagram of the THz-induced phase modulation of a probe beam imparted by an antenna array.} (a) Sketch depicting an array of THz antenna illuminated simultaneously by an incident THz pulse. In these settings, the probe pulse path is fixed, while the THz pulse path is varied via an external delay line that scans the mutual delay between the probe and THz pulses. The specific time delay corresponding to the travel time the probe pulse takes to move between consecutive antennas is $\Delta \tau$. Note that the impinging THz pulse is drawn with a tilt angle only for the sake of clarity (the incidence axis is purely orthogonal to the array plane). (b) As the time delay increases, the THz pulse path shrinks, making the THz pulse arrive at the chip at ever earlier times. Thus, the probe pulse will effectively interact with the THz near-field of an ever-increasing number of antennas as it propagates towards the end of the array (green, red, and blue solid lines moving from earlier to later time delays). The most right antenna is considered as the reference for this time frame (vertical dashed line). The total phase modulation applied to the probe pulse (black solid line) is thus the temporally-spaced superposition of all antenna contributions. Note that while the blue, red and green curve represent actual THz electric field waveforms, the black curve is more properly a representation of the THz-induced probe phase modulation and it is vertically shifted for clarity only.} 
  \label{Phase_transient_formation}
\end{figure}

The digression presented in the previous section provides an analytical description of the spectral response of the whole MZI device. However, owing to the transient nature of the reconstructed THz signal, the operation of the device (specifically, of the antenna arrays) can be effectively described in the time domain, by considering the interplay of THz near-field oscillations established in the gap of each antenna as the mutual delay between the incoming THz and probe pulses is being changed. We recall from Supplementary Fig. \ref{fig3_setup} that in our setup the probe path is kept fixed, while the mechanical delay line varies the length of the THz beam path. In particular, all measurements shown throughout the main text have been recorded by moving the delay line along its forward direction, thus shrinking the THz path. Therefore, longer time delays in the temporal axis correspond to the THz pulse that arrives earlier at the device. Let us now consider Supplementary Fig. \ref{Phase_transient_formation}a, depicting the probe beam about to cross an array of three antennas. Let us also assume that the delay line is stopped at a position that leads the probe beam to firstly encounter the THz pulse at the most right antenna (Ant. 1, blue gap), right at the time instant when the THz near-field oscillation exhibits its peak (marked as $\Delta \tau = 0$). In this case, since all antennas resonate synchronously, in order for the probe beam to be temporally aligned with the emergence of the THz peak at the last antenna, it has to cross the previous two antennas without experiencing any phase modulation, as the latter were not excited by the incoming THz electric field yet. 

Since each antenna is temporally separated from its neighbors by a time interval of $\Delta \tau$, the probe beam will encounter the THz peak at the second antenna (Ant. 2) only after the delay line has scanned an extra time interval equal to $\Delta \tau$, anticipating the arrival of the THz pulse arrival at the chip. However, contrary to the first case, after crossing Ant. 2, the probe beam will move towards Ant. 1 which has already started resonating, thus further interacting with its THz near-field, yet during a delayed evolution of the THz oscillations. Similar considerations can be made for all the subsequent antennas. Supplementary Fig. \ref{Phase_transient_formation} shows the temporal order of the emergence of the THz transients corresponding to each antenna as experienced by the probe beam. The black curve represents the resultant phase modulation transient imparted to the probe beam due to the linear superposition of all antenna contributions.
We point out that the final waveform exhibits an oscillating behavior with each local maximum corresponding to the arrival time of the probe beam at each antenna. Indeed, the case represented in this figure corresponds to the device explored in the main manuscript and it has to be ascribed to the fact that the antenna resonance is detuned from the phase-matching frequency of the array. 

\begin{figure}[htb]
 \includegraphics[width=17cm]{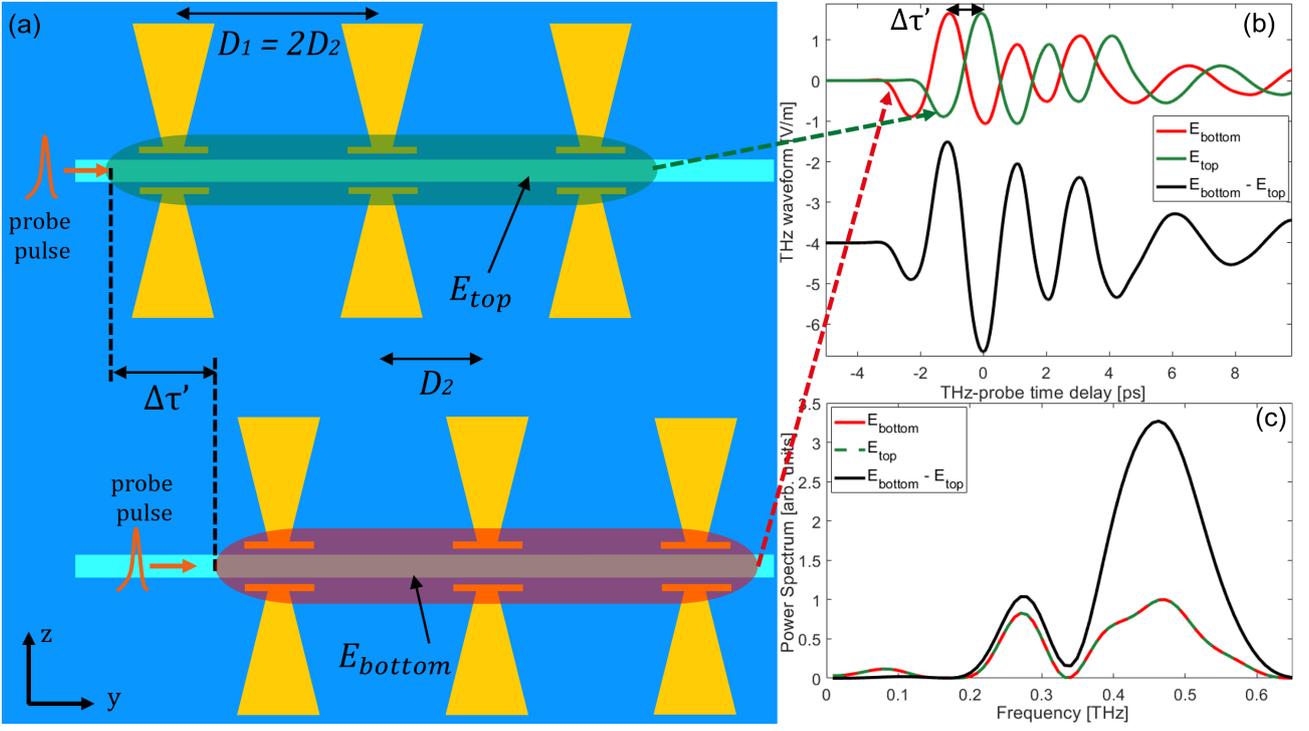} \renewcommand{\baselinestretch}{1}
 \centering
   \caption{\textbf{Time diagram of the THz-induced intensity modulation of a probe beam imparted by the interference of two displaced arrays of antennas.} (a) Sketch depicting two arrays of antennas, each running along one arm of a MZI device (not shown in the figure). Both top (red halo) and bottom (green halo) arrays generate the same THz-induced phase modulation transient (as that in Supplementary Fig. \ref{Phase_transient_formation}) that modulates each corresponding sub-probe. The different color of the halo enveloping each array indicates that the THz near-field established across the corresponding arm only differs from a phase factor. (b) The longitudinal displacement between the two arrays leads to a mutual temporal delay $\Delta \tau' =  \Delta \tau/2$ between the arrival times of the two sub-probe and in turn of the corresponding time-dependent phase modulation (red and green solid lines for the top and bottom array, respectively). This leads to a specific interference of the two sub-probes, which intensifies and converts the total phase modulation into the amplitude modulation of the output probe beam (black solid line). The latter is vertically shifted for clarity only. (c) Fourier Transforms of the curves in (b). The spectrum of the interference transient reveals a significant enhancement of the component at the phase-matching frequency of the array (i.e., $1/\Delta \tau$, black solid line) compared to the case of a single array (red and green lines). Curves are normalized to the maximum of the interference spectrum (black solid line).} 
  \label{Phase_transient_formation_double_array}
\end{figure}

Consequently, any time the probe beam is temporally aligned to the THz peak of a specific antenna, the remaining oscillations of the THz transients in neighboring antennas do not significantly contribute to its phase modulation. This fact allows us to precisely associate each waveform peak to the spatial position of a specific antenna along the array, which is the key feature enabling the beam profiling capability provided by our device, as described in the main text. The time diagram used to explain the order of interaction between the probe beam and each antenna within an array can be used to interpret the formation of the total phase modulation transient acquired by the probe beam when the two arms of the MZI interfere, as depicted in Supplementary Fig. \ref{Phase_transient_formation_double_array}. In particular, panel (a) shows the case of a device where the array on the bottom arm is displaced in the forward direction by a temporal delay equal to $\Delta \tau'$ compared to that on the top arm. By considering the entire array as a single entity that gives rise to an overall phase modulation transient as that seen in Supplementary Fig. \ref{Phase_transient_formation}b, we can draw the plot shown in Supplementary Fig. \ref{Phase_transient_formation_double_array}b. Due to the array displacement being equal to half of that between consecutive antennas (i.e., $\Delta \tau' = \Delta \tau/2$), each sub-probe in the interferometer will experience a phase modulation of mutual opposite polarity. As derived in Eq. \ref{inteference_electric}, the superposition of the electric field of the two sub-probes at the output of the interferometer results in the difference between the two individual phases. Therefore, the final phase modulation transient imparted to the recombined probe beam is the difference between the single array, which corresponds to a positive adding up of the two contributions, due to the inherent opposite polarity. It is worth noticing that the interference between the two arms preserves the time position of each peak in the final waveform, and thus, in turn, the bijective mapping with the spatial location of each single antenna along both arrays. Finally, Supplementary Fig. \ref{Phase_transient_formation_double_array}c shows the power spectra associated with the single array and double displaced cases. We note that the interference between the two arms enhances the component at the phase-matching frequency, which is the dominant oscillation cycle in the corresponding waveform. This further filtering action due to the waveforms interference improves the visibility of each waveform peak, and in turn the beam profiling capability.

%\section{Supplementary Note 6: Concurrent resonances of THz antennas and array}\label{Supp:Array_antenna_detuning}
%\begin{figure}[htb]
%\includegraphics[width=17cm]{figures/Figure_S_9_1.png} \renewcommand{\baselinestretch}{1}
% \centering
   %\caption{\textbf{Terahertz transients in antenna arrays operating at tuned and de-tuned conditions.} Simulated THz waveforms reconstructed via MZI hosting displaced double arrays with an increasing number of antennas per arm (see legend) in the case of (a) perfectly tuned and (c) detuned antenna resonance with the phase-matching frequency of the array. Curves are vertically shifted for clarity. (b) and (d) Power spectra of the temporal curves in (a) and (c) calculated via Fourier Transform. The insets in both (b) and (d) show in more detail the spectral region around the first non-null harmonic of the array (around 1.5 THz). Note that the spectrum corresponding to the single array has been magnified by 100 in both cases, to present a clear comparison.} 
 % \label{Comparison_Tuned_Detuned}
%\end{figure}
%\begin{figure}[htb]
% \includegraphics[width=17cm]{figures/Figure_S_9_2.png} \renewcommand{\baselinestretch}{1}
% \centering
%   \caption{\textbf{Amplitude peak trends of the simulated THz signals as a function of the detuning between antenna resonance and array phase-matching frequency.} Retrieved amplitude peak trends from the curves shown in Supplementary Fig. \ref{Comparison_Tuned_Detuned} for the (a) tuned and (b) detuned case.} 
%  \label{Comparison_peaks_tuned_detuned}
%\end{figure}

\newpage

\bibliography{references_MZM_TFLN}
\bibliographystyle{paper}